\documentclass[%
 aip,
 jrse,
 floatfix,
preprint,%
author-year,
]{revtex4-1}

\usepackage{graphicx}
\usepackage{dcolumn}
\usepackage{bm}
\usepackage[utf8]{inputenc}
\usepackage[T1]{fontenc}
\usepackage{mathptmx}
\usepackage{etoolbox}
\usepackage{graphicx}
\usepackage{float}
\usepackage{stfloats}
\usepackage{appendix}
\usepackage{amsmath}
\usepackage{longtable}
\usepackage{lineno}
\usepackage{array}
\usepackage{tikz}
\usepackage{mwe,amssymb,bm}
\usepackage{lipsum}
\usepackage{ulem}
\usepackage[most]{tcolorbox}

\newtcbox{\myhighlight}{on line, colframe=yellow, colback=yellow!30, sharp corners, boxrule=0pt, boxsep=0pt, left=2pt, right=2pt, top=1pt, bottom=1pt}

\newcommand\numberthis{\addtocounter{equation}{1}\tag{\theequation}}
\newcommand{\mean}[1]{\left\langle#1\right\rangle}

\newcommand\T[1]{\hat{\mathrm{#1}}}

\definecolor{color0}{RGB}{0, 191, 191}   
\definecolor{color1}{RGB}{255, 0, 0}         
\definecolor{color2}{RGB}{119, 172, 48}      
\definecolor{color3}{RGB}{0, 0, 0}           
\definecolor{color4}{RGB}{255, 0, 255}       
\definecolor{color5}{RGB}{0, 255, 255}       
\definecolor{color6}{RGB}{126, 47, 142}      
\definecolor{color7}{RGB}{0, 0, 255}         
\definecolor{color8}{RGB}{237,177, 32}      
\definecolor{MATLAB_green}{RGB}{0,255, 0}      

\newcommand{\TNBLcirc}{\raisebox{0pt}{\tikz{\draw[color0,fill = color0] (2.5mm, 0) circle (1.75pt);}}}
\newcommand{\CNBLcirc}{\raisebox{0pt}{\tikz{\draw[color1,fill = color1] (2.5mm, 0) circle (1.75pt);}}}
\newcommand{\CNBLcircopen}{\raisebox{0pt}{\tikz{\draw[color1,thick] (2.75mm, 0) circle (1.75pt);}}}
\newcommand{\blackcircopen}{\raisebox{0pt}{\tikz{\draw[black,thick] (2.75mm, 0) circle (1.75pt);}}}
\newcommand{\magentacircopen}{\raisebox{0pt}{\tikz{\draw[color4,thick] (2.75mm, 0) circle (1.75pt);}}}
\newcommand{\SBLacirc}{\raisebox{0pt}{\tikz{\draw[color2,fill = color2] (2.5mm, 0) circle (1.75pt);}}}
\newcommand{\SBLbcirc}{\raisebox{0pt}{\tikz{\draw[color3,fill = color3] (2.5mm, 0) circle (1.75pt);}}}
\newcommand{\SBLccirc}{\raisebox{0pt}{\tikz{\draw[color4,fill = color4] (2.5mm, 0) circle (1.75pt);}}}
\newcommand{\SBLdcirc}{\raisebox{0pt}{\tikz{\draw[color5,fill = color5] (2.5mm, 0) circle (1.75pt);}}}
\newcommand{\SBLdtriangle}{
    \tikz[baseline=-0.5ex] \draw[thick,yshift=-0.03cm,color5] (0,0.125) -- (0.15,0.125) -- (0.075,-0.025) -- cycle;
}
\newcommand{\SBLecirc}{\raisebox{0pt}{\tikz{\draw[color6,fill = color6] (2.5mm, 0) circle (1.75pt);}}}
\newcommand{\SBLfcirc}{\raisebox{0pt}{\tikz{\draw[color7,fill = color7] (2.5mm, 0) circle (1.75pt);}}}
\newcommand{\SBLftriangle}{
    \tikz[baseline=-0.5ex] \draw[thick,yshift=-0.03cm,color7,rotate around={90:(0.075,0.075)}] (0,0.125) -- (0.15,0.125) -- (0.075,-0.025) -- cycle;
}

\newcommand{\Iumodel}{\raisebox{2pt}{\tikz{\draw[-, color8, line width = 2.0pt](0,0) -- (5mm,0);}}}

\newcommand{\redcirc}{\raisebox{0pt}{\tikz{\draw[red,thick] (2.5mm, 0) circle (1.75pt);}}}

\newcommand{\bluetrianglemarker}{
    \tikz[baseline=-0.5ex] \draw[thin,yshift=-0.03cm,color=blue] (0,0.125) -- (0.15,0.125) -- (0.075,-0.025) -- cycle;
}
\newcommand{\trianglemarkereast}{
    \tikz[baseline=-0.5ex] \draw[thin,yshift=-0.03cm,rotate around={90:(0.075,0.075)}] (0,0.125) -- (0.15,0.125) -- (0.075,-0.025) -- cycle;
}

\newcommand{\cyandotline}{\raisebox{2pt}{\tikz{\draw[color5,dotted,line width = 2.0pt](0,0) -- (5mm,0);}}}
\newcommand{\magentadashdotline}{\raisebox{2pt}{\tikz{\draw[-,color4, dash dot,line width = 2.0pt](0,0) -- (5mm,0);}}}
\newcommand{\cyandashdotline}{\raisebox{2pt}{\tikz{\draw[-,color5, dash dot,line width = 1.0pt](0,0) -- (5mm,0);}}}
\newcommand{\blackline}{\raisebox{2pt}{\tikz{\draw[-,black,solid,line width = 1.0pt](0,0) -- (5mm,0);}}}

\newcommand{\blueline}{\raisebox{2pt}{\tikz{\draw[-,blue,solid,line width = 1.0pt](0,0) -- (5mm,0);}}}

\usepackage{hyperref}
\hypersetup{
	colorlinks=true,
	linkcolor=red,
	filecolor=magenta,    
	citecolor=blue,  
	urlcolor=cyan,
}
\allowdisplaybreaks

\makeatletter
\def\@email#1#2{%
 \endgroup
 \patchcmd{\titleblock@produce}
  {\frontmatter@RRAPformat}
  {\frontmatter@RRAPformat{\produce@RRAP{*#1\href{mailto:#2}{#2}}}\frontmatter@RRAPformat}
  {}{}
}%
\makeatother
\begin{document}

\preprint{AIP/123-QED}


\title{An extended analytical wake model and applications to yawed wind turbines in atmospheric boundary layers with different levels of stratification and veer}

\author{Ghanesh Narasimhan}
\email{naras062@umn.edu}
 \affiliation{St. Anthony Falls Lab. \& Department of Mechanical Engineering, University of Minnesota, Minneapolis, MN, USA, 55414.}
\author{Dennice F. Gayme}%
\affiliation{ 
Department of Mechanical Engineering, Johns Hopkins University, Baltimore, MD, USA, 21218.
}%

\author{Charles Meneveau}
\affiliation{ 
Department of Mechanical Engineering, Johns Hopkins University, Baltimore, MD, USA, 21218.
}%

\date{\today}

\begin{abstract}
Analytical wake models provide a computationally efficient means to predict  velocity distributions in wind turbine wakes in the atmospheric boundary layer 
(ABL). Most existing models are developed for neutral atmospheric conditions and correspondingly neglect the effects of buoyancy and Coriolis forces that lead to veer, i.e. changes in the wind direction with height. Both veer and changes in thermal stratification lead to lateral shearing of the wake behind a wind turbine, which affects the power output of downstream turbines. 
Here we develop an analytical engineering wake model for a wind turbine in yaw in ABL 
flows including Coriolis and thermal stratification effects. The model combines the new analytical representation of ABL vertical structure based on coupling Ekman and surface layer descriptions  \citep{Narasimhan_et_al_BLM_2024} with the vortex sheet-based wake model for yawed turbines \citep{bastankhah_et_al_2022}, as well as a new method to predict the wake expansion rate based on the Townsend-Perry logarithmic scaling of streamwise velocity variance. The proposed wake model's predictions show good agreement with Large Eddy Simulation (LES) results, capturing the effects of wind veer and yawing including the curled and sheared wake structures across various states of the ABL, ranging from neutrally to strongly stably stratified atmospheric conditions. The model significantly improves power loss predictions from wake interactions, especially in strongly stably stratified conditions where wind veer effects dominate.
\end{abstract}
\maketitle
   
\begin{quotation}
    \noindent \textit{The following article has been submitted to the Journal of Renewable and Sustainable Energy. After it is published, it will be found at \url{https://pubs.aip.org/aip/jrse}.}
\end{quotation}

\section{Introduction \label{sec:intro}}

Wind turbines operating in the atmospheric boundary layer (ABL) generate a wake region characterized by reduced wind speeds that affect the performance of downstream turbines. Such wake-turbine interactions can significantly reduce the overall power output of wind farms. High-fidelity numerical simulations have contributed to our understanding of wind farm wake interactions \citep{calaf_et_al_2010,stevens_et_al_JRSE_2014,Abkar_Porteagel_2015,Allaerts_Meyers_2015}. 
However, such simulations are associated with high computational costs. In many applications fast-running engineering wake models can be used instead, to approximate the velocity deficits due to turbine wakes and wake interaction, estimate average wind farm power outputs, and improve wind turbine layout during wind farm design. 
The accuracy of wake models' power predictions relies on their ability to faithfully represent the physics of wind flows within wind farms under the various atmospheric conditions in which they can be expected to operate.  The most widely used wake models \citep{stevens_meneveau_2017} represent the velocity deficit behind a wind turbine either as a top-hat distribution \citep{Jensen_1983,katic1986} or Gaussian distribution \citep{bastankhah2014,bastankhah2016}, which are axially symmetric and thus do not include effects from veer or turbine yaw. The wake deficit is then added to an assumed incoming wind profile, which can be either uniform or follows that classical Monin-Obukov Similarity Theory (MOST) scaling \citep{Monin_Obukhov_1954} that is valid in the lower portions of the ABL. 

As wind turbines become taller, interest has grown in more realistic analytical models for the incoming velocity distribution in the ABL, valid at higher elevations and across the entire ABL. Even for the relatively simple case of a conventionally neutral boundary layer (CNBL), several regimes coexist at various heights. A neutral turbulent region, with an inner surface layer where roughness is important, is separated above from a stably stratified outer flow by a thin capping inversion layer (typically  at a height of around 1-2 km) \citep{stull_1988,Liu_Liang_2010}. Above, the free-stream outer flow is in a Geostrophic balance, characterized by a flow (the geostrophic wind) in which pressure gradient and Coriolis forces are in balance. Deviations from the CNBL regime arise when the Earth's surface undergoes temperature changes due to heating or cooling of the ground. Daytime solar heating induces an unstable or convective boundary layer (CBL), where convective motions intensify turbulence within the ABL. Conversely, as the ground cools at night, a stably stratified ABL (SBL) flow develops, with much-reduced turbulence levels and lower ABL height. It is well-known \citep{Abkar_Porteagel_2015} that the thermal characteristics of the ABL greatly influence wind turbine wake properties, (e.g. wake recovery rate) and the overall performance of wind farms. Increased turbulence within the CBL flows facilitates faster recovery of the wind turbine wakes, whereas the lower turbulence within SBL flows leads to a slower recovery rate and longer wake structures. 

In addition to thermal stratification effects, the ABL is characterized by an Ekman spiral-like mean velocity distribution, induced by the Coriolis force and surface drag interactions \citep{ekman1905}. This spiral flow gives rise to a height-dependent lateral realignment of the incoming wind direction known as wind veer, which can significantly impact wind farm power output \citep{Gadde_2019}. The strength of the wind veer in the Ekman boundary layer depends on atmospheric thermal stability. Specifically, wind veer is most prominent in SBL flows, whereas increased vertical mixing resulting from convection in a CBL leads to weaker veering  \citep{Deardorff_1972, wyngaard_2010, berg_et_al_2013, Liu_Stevens_RE_2022}. 
The impact of wind veer in an SBL flow on the wake of an unyawed wind turbine was previously explored in \cite{Abkar_Porteagel_2016}, and \cite{Abkar_et_al_2018}. These studies showed that wind veer induces lateral shearing in the wake structure, resulting in a deviation from the axisymmetric wake shape assumed in traditional wake models. Consequently, wind turbines situated in ABL flows characterized by substantial stable stratification exhibit wake structures that are not only longer due to a reduced wake recovery rate but also highly sheared in the lateral direction. These findings highlight the importance of incorporating the influences of veer and thermal stratification into analytical wake models. The veer correction concept introduced in \cite{Abkar_et_al_2018} was extended to the case of 
a yawed wind turbine in a CNBL flow by \textcite{Narasimhan_et_al_2022}. Their results demonstrated that incorporating the veer correction term into the vortex sheet-based wake model  \citep{bastankhah_et_al_2022} reproduces a sheared wake structure on top of the wake curling effects due to turbine yawing. However, computation of these veer correction terms required ABL velocity profiles obtained separately from LES as inputs, rendering the wake model not fully predictive or self-consistent. 

In a recent study, \textcite{Narasimhan_et_al_BLM_2024} proposed an analytical model capable of predicting the complete vertical structure of 
thermally stratified ABL
flows, by coupling Ekman and surface layer velocity profiles in a self-consistent manner. 
The model provides analytical expressions for 
both the streamwise and spanwise velocity components, 
capturing the Ekman spiral as well as the near-surface MOST behavior across 
CNBL and SBL atmospheric conditions. 
The model also offers analytical predictions for friction velocity $(u_*)$ and cross-isobaric angle ($\alpha_0$) of the Geostrophic wind by self-consistent matching of the Ekman and surface layer velocity profiles. The model equations for predicting $u_*$ and $\alpha_0$ are commonly referred to as the Geostrophic Drag Law (GDL) \citep{Zilitinkevich_Esau_2005,Liu_et_al_2021}.
In the preliminary work reported in \textcite{Narasimhan_et_al_JPCS_2024}, the new coupled Ekman-surface layer analytical ABL model from \cite{Narasimhan_et_al_BLM_2024} was combined with \textcite{bastankhah2014}'s Gaussian wake model and used to predict wake structures 
in both neutral and stable atmospheric conditions for an unyawed wind turbine.

A more complete wake model should also include the effects of turbine yaw.  Wind farm performance for a given layout can be improved by yawing a wind turbine to deflect the wake out of the path of downstream turbines \citep{fleming2019,howland19}. Yawing a turbine can be used to redirect a wake away from downstream turbines. Yaw-induced wake deflection has been attributed to the formation of a counter-rotating vortex pair (CVP), generated by a height-dependent transverse turbine thrust force \citep{howland16, bastankhah2016, shapiro2018, shapiro2020}. The sidewash velocity from the vortex pair deflects the wake laterally, while also contributing to the curled (deforming) wake shape.
In \textcite{shapiro2018}, an analytical model for yawed turbine wakes was proposed, where the rotor area of the turbine is treated as a lifting surface that applies a height-dependent sideways force onto the fluid. 
The evaluation of the induced strength of the CVP near the turbine enabled the accurate prediction of yaw-induced wake deflection. Other vortex-based models describe the vorticity at the turbine as a distribution of multiple discrete point vortices \citep{Martinez_et_al_2020,Martinez_Tossas_2020,zong_porte_2020}, with downstream transport and diffusion modeled numerically. Following these studies, \textcite{shapiro2020} proposed a theory for the generation and downstream evolution of the CVP. Analytical predictions for the decay of the maximum vorticity and circulation strength of the vortices showed good agreement with Large Eddy Simulation (LES) data. However, this model assumed a simplified circular wake shape, neglecting the known wake deformation (curling) behavior. \textcite{bastankhah_et_al_2022} addressed this problem by proposing a vortex sheet-based wake model that predicts the curled wake shape behind yawed turbines. Within this model, the wake edge was treated as a vortex sheet, and analytical solutions for this vortex sheet were obtained using truncated power series expansions based on the decaying circulation strength estimate of the CVP from \textcite{shapiro2020}. Subsequently, \textcite{bastankhah_et_al_2022} enhanced the Gaussian wake model for the velocity deficit by incorporating the deformation caused by the vortex sheet, accurately predicting the curled wake shape and wake deflection.

In this study, we develop an extended model to predict wakes behind wind turbines (yawed and unyawed) placed in CNBL and SBL atmospheric conditions, by integrating the coupled Ekman-surface layer analytical ABL model \citep{Narasimhan_et_al_BLM_2024} with \textcite{bastankhah_et_al_2022}'s vortex sheet-based wake model. {\color{black}We also discuss a model to vary the wake expansion rate according to a given atmospheric condition based on the Townsend-Perry logarithmic scaling of streamwise velocity variance}. We then validate the extended wake model through comparisons to large-eddy simulations (LES) of unyawed and yawed wind turbines under a variety of atmospheric conditions including CNBL and SBL flows. This manuscript comprises five major sections. In section \ref{sec:wake_ABL_model}, we discuss the new integrated analytical wake model. We use LES data for certain parts of the wake model development, as well as to perform validation tests. The details of the LES are discussed in Section \S\ref{sec:LES}. The wake model performance and accuracy is demonstrated by comparing model predictions with LES, in Section \S\ref{sec:results}. Conclusions of the study are summarized in Section \S\ref{sec:conclusion}.

\section{Extended analytical wake model}\label{sec:wake_ABL_model}

In this section, we summarize the various ingredients of the extended wake model that characterizes the wake structure behind a wind turbine yawed by an angle $\beta$, operating in conventionally neutral and stable atmospheric conditions. The model integrates the vortex sheet-based wake model introduced in \textcite{bastankhah_et_al_2022}, a veer-correction term \citep{Abkar_et_al_2018,Narasimhan_et_al_2022,Narasimhan_et_al_JPCS_2024}, and the coupled Ekman-surface layer ABL wind model introduced in \textcite{Narasimhan_et_al_BLM_2024}. 
The model focuses on the mean velocity deficit $\Delta u = U(z) - u(x, y, z)$ expressed as the nominally Gaussian distribution 
\begin{align}
\frac{\Delta u}{U_h} = C(x) \exp\left[-\frac{(y-y_c)^2+(z-z_h)^2}{2\ \sigma(\theta, x)^2}\right]. \label{du_model}
\end{align}
Here, $x$, $y$, and $z$ represent the streamwise, spanwise, and wall-normal directions in the Cartesian coordinate system. The term $C(x)=\Delta u_{\text{max}}/U_h$ denotes the magnitude of the maximum velocity deficit ($\Delta u_{\text{max}}$) normalized by the ABL inflow velocity $(U_h)$ at hub height $(z_h)$. The parameter $y_c$ denotes the centroid location of the wake in the spanwise direction, and $\sigma(\theta, x)$ represents the width of the wake in different directions $\theta$ (the polar angle measured with respect to the wake's centroid location). Having a $\theta$-dependent wake width enables us to generate deformed (``curled'') shapes based on the original axisymmetric Gaussian distribution \citep{bastankhah_et_al_2022}. 
In the subsequent subsections, we discuss the various ingredients needed to fully evaluate the velocity deficit according to Eq. \eqref{du_model}.

\subsection{Maximum velocity deficit magnitude  in ABL flows}
The maximum velocity deficit magnitude $C(x)$ is modeled as \citep{bastankhah_et_al_2022}
\begin{align}
 C(x)&=
\begin{cases}
2a & x\leq x_0,\\
1-\sqrt{1-\dfrac{C_T\cos^3\beta}{2\tilde{\sigma}^2(x)/R^2}}    & x\geq x_0
\end{cases}.
\label{Cx}
\end{align}
Here, $C_T$ is the turbine's thrust coefficient, $R$ is the radius of the turbine rotor and  
\begin{align}
\tilde{\sigma}^2(x)=(k_w  x+0.4 R\sqrt{A_*})(k_w  x+0.4 R\sqrt{A_*}\cos\beta). \label{sigma_tilde}
\end{align} 
Here, $A_*$ is the ratio of the expanded stream tube area to the projected frontal area of the rotor:
\begin{align}
    A_*=\frac{1+\sqrt{1-C_T\cos^2\beta}}{2\sqrt{1-C_T \cos^2\beta}},\label{A_star} 
\end{align}   and $k_w$ denotes the wake expansion rate.

The parameter $x_0$ in Eq. \eqref{Cx}  delineates the transition point from the potential core region to the decay region. In the potential core region, the flow velocity is $U_h(1-2a)$,
where $a$ is the induction factor defined as
\begin{align}
a &= \frac{1}{2}\left(1-\sqrt{1-C_T\cos^2\beta}\right). \label{ind_factor}
\end{align}
{\color{black}As a result, from $U_h(1-2a)$, we have} $C(x \leq x_0) = 2a$ in Eq. \eqref{Cx}. For $x > x_0$, 
$C(x)$ decays due to entrainment and wake recovery (i.e. due to the increase in $\tilde{\sigma}$ with $x$ when used in Eq. \ref{Cx}).

An equation for the transition location $x_0$ can be derived by evaluating the decaying expression for $C(x)$ at $x=x_0$ and equating it to the velocity deficit magnitude  in the potential core region ($C(x\leq x_0)=2a$), which yields:
\begin{align*}
x_0 &= \frac{1}{5 k_w} \Biggl(\sqrt{R^2 A_*(1-\cos\beta)^2 +25 \frac{R^2}{2}\frac{C_T\cos^3\beta}{(1-(1-2a)^2)}}
+R\sqrt{A_*} (1-\cos \beta)\Biggr)-\frac{2}{5k_w}R\sqrt{A_*}.\numberthis\label{x0_transition}
\end{align*}
Qualitatively, it is evident from Eq. \eqref{x0_transition} that $x_0$ is inversely proportional to  $k_w$. As will be included later in the model, stable stratification leads to smaller $k_w$ and hence to more elongated, stronger, wakes.

\subsection{Polar angle dependent wake width $\sigma(\theta,x)$}
The wake width $\sigma(\theta,x)$ is calculated by incorporating the linear wake growth from the Jensen model \citep{Jensen_1983} and the angle-dependent wake shape function \citep{bastankhah_et_al_2022}
\begin{align*}
\sigma(\theta,x) = k_w \, x + 0.4 \, \xi(\theta,x), \numberthis\label{sigma_width}
\end{align*}
where
\begin{align*}
    \theta=\tan^{-1}\left(\frac{z-z_h}{y-y_c}\right)\numberthis\label{polar_angle}
\end{align*}
is the polar angle.
The wake shape function 
\begin{align}
\xi(\theta,x) =  \xi_0(\theta) \, \hat{\xi}(\theta,\hat{t}),\label{wake_shape_function}
\end{align}
where $\xi_0(\theta)=\xi(\theta,0)$ represents the initial shape of the wake and $\hat{\xi}(\theta,\hat{t})$ is the angle-dependent dimensionless vortex-sheet radial position, which is a function of a dimensionless time-like auxiliary variable $\hat{t}$.
The initial shape depends on the angle $\theta$ and takes the form of an ellipse when the rotor disk is yawed, the governing expression is given by
\begin{align}\xi_0(\theta)=R\sqrt{A_*}\frac{|\cos\beta|}{\sqrt{1-\sin^2\beta\sin^2\theta}} \label{xi0_theta}.
\end{align}
The wake-curling effects are modeled using an empirical expression for the non-dimensional vortex sheet location $\hat{\xi}(\theta,\hat{t})$ \citep{bastankhah_et_al_2022}:
\begin{align*}
\hat{\xi}(\theta,\hat{t})&=1-\alpha\biggl[
\frac{1}{2}\tanh\left(\frac{\hat{t}^{2}}{4\alpha}\right)\cos2\theta
-\frac{1}{4}\tanh\left(\frac{\hat{t}^{3}}{8\alpha}\right)\cos(3\theta)\\
&\mspace{50mu}
-\frac{5}{48}\tanh\left(\frac{\hat{t}^{4}}{16\alpha}\right)\cos(2\theta)
+\frac{7}{48}\tanh\left(\frac{\hat{t}^{4}}{16\alpha}\right)\cos(4\theta)  \biggr],\numberthis\label{xi_emp}
\end{align*}
where the model parameter $\alpha=1.263$ is used from fits to data. The dimensionless time $\hat{t}$ is:
\begin{align*}
    \hat{t}(x,z)&\approx-1.44\frac{U_h}{u_*}\frac{R}{R\sqrt{A_*}}C_T\cos^2\beta\sin\beta 
 \,\, \left[1-\exp\left(-0.35\frac{u_*}{U(z)}\frac{x}{R}\right)\right].\numberthis\label{that}
\end{align*}
Expressions \eqref{xi0_theta}, \eqref{xi_emp}, and \eqref{that}
fully specify the expression for $\sigma(\theta,x)$ in Eq. \eqref{sigma_width} for a given $k_w$.

\subsection{Wake expansion coefficient $k_w$}\label{sec:kw}

A common estimate for the wake expansion coefficient uses the ratio of transverse mixing velocity (proportional to friction velocity $u_*$ in the ABL) and hub-height advection velocity $U_h$ 
\citep{Tennekes_Lumley_1972,shapiroetal2019}. In \textcite{bastankhah_et_al_2022} $k_w=0.6 u_*/U_h$ was used, where the proportionality factor 0.6 was determined empirically from LES data for truly neutral atmospheric conditions. However, the proportionally varies for different atmospheric stability conditions. This constant can be obtained by including dependence on the streamwise turbulent intensity $I_u=\langle\overline{u^\prime u^\prime}\rangle^{1/2}/U_h$ of the incoming ABL flow. Here, $\langle\overline{u^\prime u^\prime}\rangle$
represents the time and horizontally averaged streamwise velocity variance. Previous studies \citep{Niayifar_2016,Zhan_et_al_2020,bastankhah_et_al_2022} 
have suggested a linear relationship, $k_w=\alpha^\prime I_u$,
where $\alpha^\prime$ is an empirical constant.

Here we compute $I_u$ using the logarithmic scaling of velocity variance   \citep{Townsend_1976,PerryChong_1982,marusic_kunkel_2003, hultmark_et_al_2012,marusic_et_al_2013,meneveau_marusic_2013}: 
\begin{align}
\frac{\langle\overline{u^\prime u^\prime}\rangle}{u_*^2}=-A_1 \ln\frac{z}{h} + B_1,\label{upup_var}
\end{align}
where $h$ is the boundary layer height and $A_1$ is called the Townsend-Perry constant (a value $A_1=1.25$ has been found to represent the data well \citep{hultmark_et_al_2012,marusic_et_al_2013,meneveau_marusic_2013,stevens_wilczek_meneveau_2014}), while $B_1$ is flow dependent.
Using Eq.\eqref{upup_var} and $k_w=\alpha^\prime I_u$, we can 
rewrite $I_u$ as 
\begin{align}
I_u&=\left[A_1 \ln\frac{h}{z} + B_1\right]^{1/2}\frac{u_*}{U_h}.\label{Iu_model}
\end{align}
and express the wake expansion coefficient as 
\begin{align*}
k_w=\alpha^\prime \left[A_1 \ln\frac{h}{z} + B_1\right]^{1/2}\frac{u_*}{U_h}\numberthis\label{kw_model_high_Iu}.
\end{align*}
We set $B_1=0.6$ based on relevant   LES data for Truly Neutral Boundary Layer (TNBL, pressure gradient driven), CNBL, and SBL flows to be discussed later in the paper in  \S\ref{sec:kw_turb_intensity}. 

Based on LES of wind turbines in TNBL flows where the turbulence intensity at the turbine hub height ranged from $0.06$ to $0.15$, \textcite{Niayifar_2016} determined the constant $\alpha^\prime=0.38$. Field measurement data from e.g., \textcite{fuertes_et_al_2018} and \textcite{brugger_et_al_2019} suggested $\alpha^\prime=0.3-0.35$. In our current study, consistent with these previous findings, we use $\alpha^\prime=0.33$, which leads to predictions of $C(x)$ from Eq. \eqref{Cx} that are in good agreement with  LES data (see section \S\ref{sec:Cx}).

However, as noted in \textcite{vahidi_porte-agel_2022}, the value of $\alpha^\prime$ lying within the range between 0.3-0.35 holds only for the boundary layer flows with relatively high turbulence intensities. When turbulence intensity  is lower such as in SBL flows, the wake growth rate depends on the turbulence generated at the mixing layers that form at the interface between the wake region and incoming wind.   Typical wake spreading rates in these types of 
flows are lower, with values ranging between 0.023 and 0.043 \citep{vahidi_porte-agel_2022}. Based on these findings, 
we propose a model that smoothly merges the larger of the two options according to  
\begin{align*}
k_w=\left[0.021^n + \left(0.33 \left[1.25 \ln\frac{h}{z_h} + 0.6\right]^{1/2}\frac{u_*}{U_h} 
\right)^n\right]^{1/n}. \numberthis\label{kw_CNBL_SBL_model}
\end{align*}
Using $n=6$ leads to a good match between  LES data and model predictions for $C(x)$ given by Eq. \eqref{Cx}. 
Needed values of $u_*$, $h$, $U_h$ are obtained self-consistently from the analytical ABL model \citep{Narasimhan_et_al_BLM_2024} discussed in Section \S\ref{sec:ABL}.

\subsection{Wake displacement}

The displacement of the wake center position due to yawing and wind veer is modeled according to
\begin{align}
y_c(x,z)=\hat{y}_c(\hat{t})R\sqrt{A_*}+y_{c,\text{veer}}.\label{yc_xz}
\end{align}
Here, $\hat{y}_c({\hat{t}})$ is a dimensionless form of wake deflection \citep{bastankhah_et_al_2022} expressed as
\begin{align*}
\hat{y}_c(\hat{t})&=\frac{(\pi-1)|\hat{t}|^3+2\sqrt{3}\pi^2\hat{t}^2+48(\pi-1)^2|\hat{t}|}{2\pi(\pi-1)\hat{t}^2+4\sqrt{3}\pi^2|\hat{t}|+96(\pi-1)^2}\text{sgn}(\hat{t}) -\frac{2}{\pi}\frac{\hat{t}}{[(z+z_h)/R\sqrt{A_*}]^2-1}.\numberthis\label{yhatc}
\end{align*}
The first term in Eq. \eqref{yhatc} models the deflection of the wake due to the sidewash velocity from the counter-rotating vortices that form due to yawing a wind turbine. The second term in Eq. \eqref{yhatc} represents the spanwise deflection of the wake due to the ground effect modeled through the addition of image vortices. The dimensionless timescale in Eq. \eqref{yhatc} is given by Eq. \eqref{that}. Furthermore, in Eq. \eqref{yc_xz}, the term  $y_{c,\text{veer}}$ represents the wake deflection due to wind veer. Following previous studies \citep{Abkar_et_al_2018, Narasimhan_et_al_2022,Narasimhan_et_al_JPCS_2024}, 
it is modeled as 
\begin{align}
y_{c,\text{veer}}=\frac{x}{U(z)}V(z),\label{yc_veer}
\end{align}
where $x$ is the streamwise distance behind the wind turbine and $U(z), V(z)$ are the incoming ABL flow's streamwise and spanwise velocity components, respectively as a function of height $z$. In the following section, we discuss an analytical model for specifying the ABL velocities in Eq. \eqref{yc_veer} in a self-consistent manner without relying on external inputs that were introduced in \cite{Narasimhan_et_al_BLM_2024}.

 \subsection{Coupled Ekman-surface layer ABL  model}\label{sec:ABL}

To describe the mean velocity distributions $U(z)$ and $V(z)$ as function of height analytically, \textcite{Narasimhan_et_al_BLM_2024} proposed to use Ekman and surface layer representations matched at some height $z_m$ (20\% of the boundary layer height $h$ was chosen).  
The coordinate system is chosen such that the stress at the ground surface is in the $x$ direction. 
The velocity profiles can be written as
\begin{align*}
 \frac{U(\hat{\xi})}{u_*}&= 
\begin{cases}
 -g^\prime(\hat{\xi})\left[1-\dfrac{\hat{\xi}}{\hat{h}}\right]^{3/2}+\dfrac{3g(\hat{\xi})}{2\hat{h}}\sqrt{1-\dfrac{\hat{\xi}}{\hat{h}}}  +  \dfrac{U_g}{u_*},
& \, \hat{\xi}\geq 0.2 \, \hat{h}\\ 
&\\
\mspace{20mu}\dfrac{1}{\kappa}\ln\left(\dfrac{\hat{\xi}}{\hat{\xi}_0}\right)+(5\mu+0.3\mu_N)(\hat{\xi}-\hat{\xi}_0)\mspace{28mu}, &\ \hat{\xi}\leq 0.2 \, \hat{h}
\end{cases},\numberthis\label{U_vel}\\
\frac{V(\hat{\xi})}{u_*}&=  \dfrac{g(\hat{\xi})g^\prime(\hat{\xi})}{{[1-g(\hat{\xi})^2]^{1/2}}}
\left[1-\frac{\hat{\xi}}{\hat{h}}\right]^{3/2}+\frac{3}{2\hat{h}}\sqrt{1-g(\hat{\xi})^2}\left[1-\frac{\hat{\xi}}{\hat{h}}\right]^{1/2} + \frac{V_g}{u_*}  .\numberthis\label{V_vel}
\end{align*} 

In these equations, 
$\hat{\xi}=z f_c/u_*$, $\hat{\xi}_0=z_0 f_c/u_*$, and $\hat{h}=h f_c/u_*$ are the dimensionless heights, surface roughness and ABL heights, respectively.  $f_c=2\Omega\sin\phi$ is the Coriolis frequency, where $\Omega=7.27\times 10^{-5}$ 1/s is the Earth's rotation rate and $\phi$ represents the latitude. In addition, $\kappa=0.41$ is the Von-K\'arm\'an constant, and $U_g=G\cos\alpha_0$, $V_g=G\sin \alpha_0$ are the Geostrophic wind velocity components that depend on the Geostrophic wind magnitude, $G=\sqrt{U_g^2+V_g^2}$,  and cross-isobaric angle, $\alpha_0=\tan^{-1}(V_g/U_g)$. Also included in the MOST portion of the model are the Monin-Kazanski stability parameter $\mu = u_*/\kappa f_c L_s$ and the Zilitinkevich number $\mu_N = N_\infty/f_c$. These account for the effects of surface cooling and the thermal stratification strength of the free stream, respectively. Here, $L_s = -u_*^3/[\kappa (g/\Theta_0) Q_0]$ is the Obukhov length, where $g = 9.81  \text{m/s}^2$ is the gravitational acceleration, and $\Theta_0$ is the reference potential temperature. The quantity $N_\infty = \sqrt{(g/\Theta_0) \gamma_\Theta}$ is the Brunt-V\"ais\"al\"a frequency of the free geostrophic flow, with $\gamma_\Theta$ denoting the vertical gradient of the potential temperature. For a quasi-steady ABL flow, the surface cooling flux can be modeled as $Q_0 = C_r h$, where $C_r$ is the cooling rate of the surface potential temperature \citep{Narasimhan_et_al_BLM_2024}.  As written in Eqs. \eqref{U_vel} \& \eqref{V_vel}, the analytical expressions describe velocity profiles for a latitude $\phi$ located in the northern hemisphere (in the southern hemisphere, a simple change in sign for spanwise velocity profile $V(z)$ and Geostrophic wind speed component $V_g$ is required). 

The function $g(\hat{\xi})$ serves to ensure consistent behavior of the Reynolds stress profiles when approaching the upper parts of the boundary layer, when $\hat{\xi}\to \hat{h}$. With its derivative, they are given by \textcite{Narasimhan_et_al_BLM_2024}:  
\begin{equation}
   g(\hat{\xi}) = c_g\left(1-e^{-{\hat{\xi}}/{\Gamma \hat{h}}}\right),\qquad
   g^\prime(\hat{\xi}) = \left({c_g}/{\Gamma \hat{h}}\right)e^{-\hat{\xi}/\Gamma\hat{h}},    \refstepcounter{equation}\tag{\theequation a,b}\label{eqn:avoidnumbers1}
\end{equation}
where, $c_g=1.43, \ \Gamma=0.83$ are model parameters obtained by fitting LES data \citep{Narasimhan_et_al_BLM_2024}. 

The $U(\hat{\xi})$ profile, as described in Eq. \eqref{U_vel}, is composed of two layers: the outer and inner layers. The velocity in the outer layer follows the classical $3/2$ power-law profile for the total stress, $\T{T} = \sqrt{\T{T}_{xz}^2 + \T{T}_{yz}^2} = [1 - \hat{\xi}/\hat{h}]^{3/2}$ \citep{Nieuwstadt_1984}, where $\T{T}_{xz}$ and $\T{T}_{yz}$ are the turbulent shear stress components in the streamwise and spanwise directions, respectively. The velocity in the inner layer is defined by the  Monin-Obukhov Similarity Theory (MOST) profile, capturing the characteristics of the Atmospheric Surface Layer (ASL). These two layers are connected at a matching height $\hat{\xi}_m=0.2 \,\hat{h}$. The boundary layer height $\hat{h}$ is modeled according to  \citep{Zilitinkevich_Esau_2005}:
\begin{align}
    \frac{1}{\hat{h}^2}=\frac{1}{C_{TN}^2}+\frac{\mu_N}{C_{CN}^2} +\frac{\mu}{C_{NS}^2}. \label{h_hat_model}
\end{align}
The model constants $C_{TN} = 0.5$ and $C_{CN} = 1.6$ are determined by fitting the modeled $\hat{h}$ with height data from the LES of CNBL flows in \cite{Liu_et_al_2021}. Similarly, the constant $C_{NS} = 0.78$ is obtained by fitting the expression for ABL height with measurements from the LES data of SBL flows described in \cite{Narasimhan_et_al_BLM_2024}.

Furthermore, for a given Geostrophic wind velocity magnitude $G$, the components $U_g$, $V_g$, and $u_*$ are determined using the new Geostrophic Drag Law (GDL) model introduced in \cite{Narasimhan_et_al_BLM_2024}:
\begin{equation}
   \kappa U_g/u_*=\ln\left(u_*/f_c z_0\right)-A,\qquad
   \kappa V_g/u_*=- B,
\refstepcounter{equation}\tag{\theequation a,b}\label{eqn:GDL}
\end{equation}
where   $A$ and $B$ are given by
\begin{align*}
A&=-\ln(0.2 \hat{h})-\kappa\biggl[(5\mu+0.3\mu_N)(0.2 \hat{h}-\hat{\xi}_0)+g^\prime(0.2 \hat{h}) \, 0.8 ^{3/2}-g(0.2 \hat{h})\,({3}/{2\hat{h}}) \, {0.8^{1/2}}\biggr],\numberthis\label{A_model_GDL}\\
B&={3\kappa}/{2\hat{h}}.\numberthis\label{B_model_GDL}
\end{align*}
\cite{Narasimhan_et_al_BLM_2024} outlines an iterative method for solving the ABL height equation and GDL equations together to determine the dimensional values of $h$, $u_*$, $U_g$, and $V_g$. These dimensional quantities are then employed to calculate the dimensionless parameters $\hat{\xi}$, $\hat{\xi}_0$, and $\hat{h}$. Substituting these results into Eqs. \eqref{U_vel} and \eqref{V_vel} yields the ABL velocity components $U(z)$ and $V(z)$. These analytical velocity profiles are subsequently used to compute the veer-correction term given by Eq. \eqref{yc_veer}.   

In summary, the enhanced version of the wake model is developed by integrating the Gaussian wake model from Eq. \eqref{du_model} with the ABL model expressions provided in Eqs. \eqref{U_vel} and \eqref{V_vel}. The coupling between the models is achieved through the veer-correction term defined in Eq. \eqref{yc_veer}. This improved wake model effectively predicts diverse wake structures across a range of atmospheric conditions, encompassing both CNBL and SBL regimes. Additionally, the model incorporates predictions for $u_*$, $U_g$, and $V_g$ using the new GDL model, represented by Eqs. \eqref{eqn:GDL}, \eqref{A_model_GDL}, and \eqref{B_model_GDL}. Conversely, if the magnitude of the geostrophic wind $G$ is given, the GDL can be used to find the angle $\alpha_0$ between the geostrophic wind direction and the surface stress as well as the friction velocity. 
The model also includes the ABL height $h$ from Eq. \eqref{h_hat_model} and the surface cooling flux $Q_0$ that is either prescribed or related to the cooling rate $C_r$ according to $Q_0 = C_r h$. The model is entirely self-contained, eliminating the need for external inputs and ensuring a comprehensive and self-consistent predictive framework. We also note another recent model by \cite{Shen_Liu_Lu_Stevens_2024} who similarly propose analytical velocity profile expressions tailored for SBL flows including GDL coefficients, while displaying smooth behavior of the profiles at the top layer limit.

\section{Large Eddy Simulation (LES) of a wind turbine in ABL flows}\label{sec:LES}

In this section, we describe the LES  of a wind turbine placed in ABL flows. The data generated from LES serves as validation for the enhanced extended wake model discussed in the preceding section \S\ref{sec:wake_ABL_model}. We employ the LESGO code, an open-source \cite{LESGO} solver developed specifically for simulating ABL flows \citep{Albertson_1999, bouzeid2005}. 
The LESGO code solves the filtered Navier-Stokes equations incorporating a buoyancy force term approximated through the Boussinesq approximation, along with the scalar potential temperature transport equation. The code has been used in a variety of prior  LES studies \citep{bouzeid2005, calaf_et_al_2010, Calaf_et_al_2011, stevens_et_al_JRSE_2014, stevens2014, martinez_et_al_2015, shapiro_et_al_AD_2019, shapiro2018, shapiro2020}.  Subgrid fluxes are modeled using a Smagorinsky eddy viscosity model with the coefficient  dynamically determined using the Lagrangian dynamic scale-dependent model \citep{bouzeid2005}. The subgrid-scale heat flux parameterization uses a prescribed subgrid-scale Prandtl number of unity. The solver adopts the concurrent-precursor method \citep{stevens2014} to generate ABL inflow for the computational domain, which includes a wind turbine represented here using the actuator disk (ADM) model \citep{Narasimhan_et_al_2022}. To mitigate the impact of streamwise periodicity, a shifted periodic boundary condition is implemented \citep{muntersetal2016}.  Details about the governing equations, boundary conditions and subgrid modeling are described in \cite{Narasimhan_et_al_BLM_2024}.  
 
For CNBL and SBL flows, the imposed Geostrophic wind direction, characterized by $\alpha$ (with $U_g=G\cos\alpha$ and $V_g=G\sin\alpha$), is regulated by a Proportional-Integral (PI) controller. This controller is designed to enforce a desired mean velocity orientation at a specific height \citep{sescu2014, Narasimhan_et_al_2022}, which is chosen to be the wind turbine hub height ($z_h$)  in this study. On the other hand, the TNBL flow is driven by an applied mean streamwise pressure gradient without the influence of buoyancy and Coriolis forces \citep{Narasimhan_et_al_2022}. 

The influence of atmospheric stability is integrated into the bottom boundary condition by computing surface momentum fluxes ($\tau_{13,w},\tau_{23,w}$) using the classical MOST expression, while a stress-free boundary condition is prescribed at the upper boundary. To mitigate the effects of gravity waves induced by thermal stratification within the computational domain, a sponge layer, commonly referred to as a Rayleigh damping layer, is introduced at the upper boundary, see e.g. \cite{allaerts_meyers_2017, Durran_Klemp_1983}. This wave-absorbing layer spans 500 meters from the upper boundary. Within this layer, a cosine-profiled body force is applied to the damping coefficient to minimize the reflection of gravity waves.

\subsection{LES setup for wind turbine in ABL flows}\label{sim_setup_section}
The LES of CNBL and SBL flows are conducted within a computational domain of dimensions $L_x\times L_y\times L_z=3.75 \ \text{km} \times 1.5 \ \text{km} \times 2\ \text{km}$. The discretization is performed in the streamwise, spanwise, and wall-normal directions using a grid of $N_x\times N_y\times N_z=360\times 144\times 432$ points. For the TNBL case, the height of the domain is set to $L_z=1$ km which is resolved using $N_z=216$ grid points. This yields a grid resolution of $\Delta x\times\Delta y\times \Delta z=10.4 \ \text{m}\times 10.4 \ \text{m} \times 4.6 \ \text{m}$ for all the ABL cases. Our grid resolution in the horizontal direction is 10 m, which is similar to the 9 m grid resolution utilized in \textcite{Gadde_Stevens_2021}.  \cite{basu_avraham_2017} recommend that MOST-based wall-modeled LES studies should use a grid size that is at a distance of at least 50$z_0$ from the surface, ensuring this grid point lies within the ASL. Hence, for $z_0=0.1$ m, a vertical grid spacing of $\Delta z\approx 5$ m is employed. The Geostrophic wind magnitude is set to $G=15$ m/s, Coriolis frequency is set to $f_c=10^{-4}$ s$^{-1}$, the surface roughness height is $z_0=0.1$ m, and the Subgrid-Scale (SGS) Prandtl number is $Pr^{\text{SGS}}=1$. For the TNBL flow, the driving mean pressure gradient force in the streamwise direction is set to a constant value \citep{Narasimhan_et_al_2022}.

\begin{figure}[H]
	\begin{center}
\includegraphics[scale=.9]{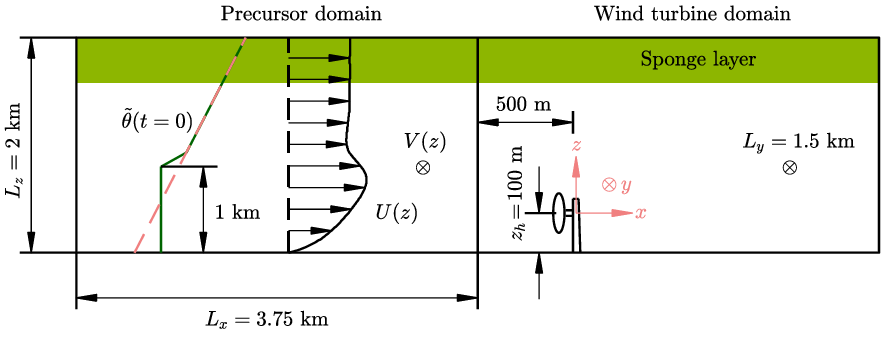}
	\end{center}
	\vspace*{0mm}
	\caption{Schematic of the LES setup for simulating a wind turbine in conventionally neutral (CNBL) and stably stratified (SBL) atmospheric flows. } \label{schematic}
\end{figure} 

In simulations of TNBL, CNBL, and SBL flows the velocity fields are initialized with a log-law velocity profile. This profile is augmented with zero-mean white noise within the first 100 m from the surface to induce turbulence and initiate the flow. The specific configurations of the initial potential temperature profile for the CNBL and SBL simulations are detailed in the following text.
\begin{table*}
\begin{ruledtabular}
\caption{\label{tab:LESGO} Summary of LES of CNBL and SBL scenarios, presenting corresponding values for the stability parameter $\mu$, friction Rossby number $Ro$, cooling rate $C_r$ (in K/hr), ABL height $h$ (in meters), friction velocity $u_*$ (in meters per second), and cross-isobaric angle $\alpha_0^\circ$. All simulations maintain consistent parameters: wind speed $G=15$ m/s, surface roughness $z_0=0.1$ m, potential temperature $\Theta_0=265$ K. Additional constants include the Coriolis frequency $f_c=10^{-4}$ s$^{-1}$ and the free-stream Brunt-Väisälä frequency $N_\infty=6.1\times 10^{-3}$ s$^{-1}$, resulting in a Zilitinkevich number $\mu_N=61$.}
\begin{tabular}{ccccccc}
            Case & $\mu$ & $Ro$ &$C_r$ (K/hr)& $h$ (m)&  $u_*$ (m/s)&  $\alpha_0^{\circ}$\\
		&&&&&&\\ 
            \hline
            &&&&&&\\
            TNBL& - &  - & 0 & 1000& 0.94 & 0\\
		CNBL& 0 &  $6.02\times 10^4$ & 0 & 1157& 0.60 &21\\
            SBL-1& 5.62 & $5.93\times 10^4$&-0.03 & 1032&  0.59 &24\\
            SBL-2& 20.59 & $5.25\times 10^4$&-0.125 & 662&  0.53 &28\\
            SBL-3& 39.84 & $4.58\times 10^4$&-0.25 & 463& 0.46 &32\\
            SBL-4& 59.25 & $4.12\times 10^4$&-0.375 & 361&0.41 &35\\
            SBL-5& 78.35 & $3.84\times 10^4$&-0.5 & 306&0.38 &38\\
            SBL-6& 148.49& $3.39\times 10^4$&-1 & 218& 0.34 &41\\
            &&&&&
\end{tabular}
\end{ruledtabular}
\end{table*}

The LES of a CNBL flow is initiated with an initial linear potential temperature profile  $\Theta(z)=\Theta_0+\gamma_\Theta z$, where $\Theta_0=265$ K and $\gamma_\Theta=0.001$ K/m. The dashed orange line in the schematic Fig. \ref{schematic} represents this linear initial potential temperature profile. The CNBL flow is obtained by applying a zero heat flux condition at the bottom boundary.
Under the influence of this insulating boundary, the simulation progresses to a quasi-stationary state, during which the boundary layer height grows to 1157 m. The resulting temperature profile exhibits a capping inversion layer at the ABL height, distinguishing the neutral boundary layer region from the stably stratified Geostrophic region. The green line in the schematic in Fig. \ref{schematic} represents this quasi-steady CNBL potential temperature profile. At this stage, the potential temperature within the CNBL region remains approximately 265.58 K. For the LES of SBL flows, we initialize the potential temperature using the CNBL's quasi-steady potential temperature profile. We then decrease the magnitude of the surface potential temperature $\tilde{\theta}_s$  by applying different cooling rates $C_r=[-0.03,-0.125,-0.25,-0.375,-0.5,-1]$ K/hr to induce various levels of stable stratification. 

Once the simulations reach a quasi-steady state, we place the wind turbine in the main domain of interest downstream of the precursor and continue the concurrent precursor simulation. A schematic representation of this setup is shown in Fig. \ref{schematic}. The wind turbine is placed 500 m from the inlet of the concurrent domain. The diameter $(D)$ and hub height $(z_h)$ of the turbine are both set to 100 m. The thrust coefficient $(C_T^\prime)$ of the wind turbine is set to 1.33. 
The simulations are performed for three different wind turbine yaw angles $\beta=0,20^\circ,-20^\circ$.   
We then perform time averaging in both the precursor and main domains obtaining the three-dimensional mean wind velocities with and without the wind turbine.
Time averaging is done over a 10-12 hour window, which we observe to be long enough for the flow to be quasi-steady with no appreciable effects from inertial oscillations. The dimensional ABL height $h$ is determined by a least-square-error minimization method for the root mean square difference between the normalized LES stress $\T{T}=\mathrm{T}(z)/u_*^2$ and the model expression $(1-z/h)^{3/2}$ \citep{Narasimhan_et_al_BLM_2024}, in a range between  $0<z<h$. The friction velocity $u_*$ is obtained from the time ($\overline{\cdot}$) and spatially ($\mean{\cdot}$) averaged shear stress components 
$\left(\mean{\overline{\tau}_{13,w}},\mean{\overline{\tau}_{23,w}}\right)$ defined as
$u_*=\left[{\mean{\overline{\tau}_{13,w}}^2+\mean{\overline{\tau}_{23,w}}^2}\right]^{1/4}$, where $\mean{\overline{\tau}_{13,w}}$, $\mean{\overline{\tau}_{23,w}}$ are obtained from the wall model used in the LES. 

The values for $h, u_*, \alpha_0$ obtained from the LES are provided in Table \ref{tab:LESGO}. Given that the PI controller ensures a streamwise aligned mean flow at $z=100$ m, the reported $\alpha_0$ values in Table \ref{tab:LESGO} are derived by geometrically rotating the mean velocity profiles. This rotation is performed in a manner that eliminates wind veer at the first grid point, aligning with the coordinate system used in the derivation of the GDL model. In this system, there is no wind veer within the ASL region, maintaining consistency with the model derivation.

\subsection{Effects of wind veer and yawing on wind turbine wakes}\label{wake_LES}

\begin{figure}[H]
    \centering
    \includegraphics[scale=0.75]{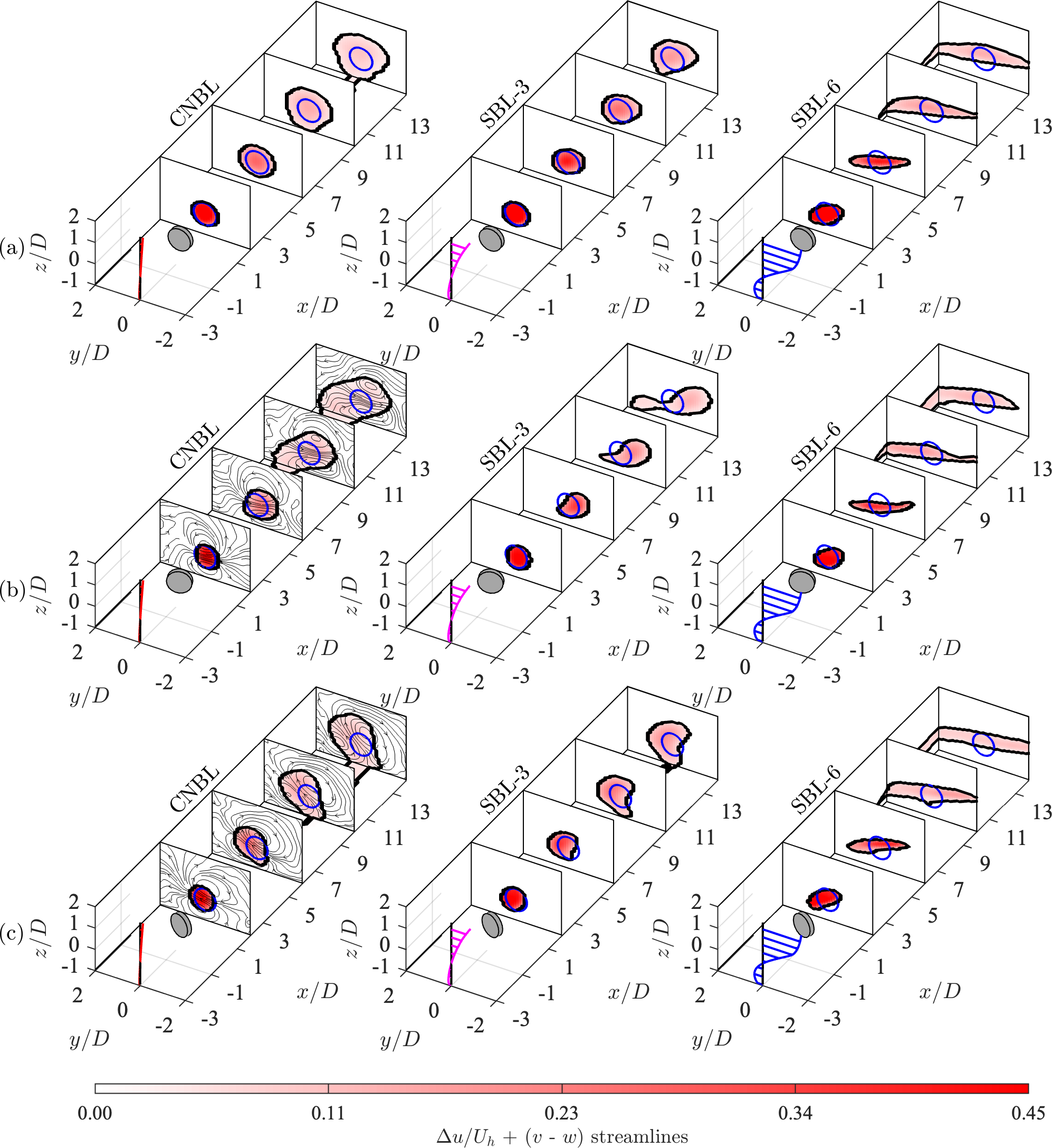}
    \caption{Wake structures downstream of a wind turbine with yaw angles (a) $\beta=0^\circ$, (b) $\beta=20^\circ$, (c) $\beta=-20^\circ$ and placed in CNBL, SBL-3 and SBL-6 ABL flows for each yaw angle. }
    \label{wakes_ABL}
\end{figure}

This section examines the combined effects of wind veer and turbine yaw on the wake structure behind a wind turbine. Fig. \ref{wakes_ABL} presents wake contour slices for a turbine in CNBL, SBL-3, and SBL-6 flows at yaw angles of $\beta=0^\circ$ (Fig. \ref{wakes_ABL}(a)), $\beta=20^\circ$ (Fig. \ref{wakes_ABL}(b)), and $\beta=-20^\circ$ (Fig. \ref{wakes_ABL}(c)).
The strength of the wind veer increases from CNBL to SBL-6 flow as shown by the vertical profile of the spanwise wind velocity plotted in front of the turbine.

The influence of only wind veer on the wake structure behind an unyawed turbine is illustrated in Fig. \ref{wakes_ABL}(a). In the CNBL flow, where the wind veer is relatively weak compared to the SBL-3 and SBL-6 cases, the wake behind the turbine exhibits a Gaussian structure that is symmetric about the turbine hub. In contrast, the wakes in the SBL-3 and SBL-6 cases are sheared, reflecting the stronger wind veer in these flows.

The combined effects of yaw and veer are illustrated in Figs. \ref{wakes_ABL}(b) and (c). Under CNBL conditions, yaw has a stronger influence on the wake than wind veer. Positive or negative yawing of the wind turbine creates a curled wake structure, deflecting the wake in the negative or positive spanwise direction, respectively. This wake curling and deflection is caused by the formation of counter-rotating vortex pairs (CVP) behind the yawed turbine \citep{howland16,shapiro2020}. The CVP flow structure is visualized by plotting the $v$-$w$ streamlines behind the yawed wind turbine under the CNBL flow scenario at various downstream distances. Notice the side wash velocity is along the negative spanwise direction in Fig. \ref{wakes_ABL}(b) (positive yaw) while it is along the positive spanwise direction in Fig. \ref{wakes_ABL}(c) (negative yaw).

In SBL-3 flow, positive yaw enhances the wake deflection, amplifying the shearing effect caused by wind veer. Conversely, with negative yaw, the wake structure is shaped by the opposing effects of yaw- and veer-induced deflections: yaw shifts the wake in the positive spanwise direction, while veer deflects it in the negative spanwise direction. In SBL-6 flow, however, the stronger wind veer dominates, overshadowing the yaw effects and producing a sheared wake structure regardless of the yaw direction.

\section{Comparisons between LES and analytical model predictions}\label{sec:results}

This section demonstrates the extended analytical wake model's effectiveness (\S\ref{sec:wake_ABL_model}) by comparing its predictions to LES results. In \S\ref{sec:kw_turb_intensity}, we compare turbulence intensity from Eq. \eqref{Iu_model} with LES data and examine wake expansion rate variations using Eq. \eqref{kw_CNBL_SBL_model} under CNBL and SBL conditions. \S\ref{sec:Cx} compares the decay of maximum velocity deficit $C(x)=\Delta u_{\text{max}}/U_h$ from Eq. \eqref{Cx} with LES results. Velocity deficit predictions are evaluated in \S\ref{sec:wake_model_vs_LES}. Finally, \S\ref{sec:power_wake_loss} compares power loss predictions from the extended wake model and Gaussian models \citep{bastankhah2014,bastankhah_et_al_2022} for a hypothetical wind turbine placed downstream of an unyawed or yawed turbine under CNBL and SBL atmospheric conditions.

 \subsection{Stability-dependent wake growth rate in ABL flows}\label{sec:kw_turb_intensity}

\begin{figure}[H]
    \centering
\includegraphics[scale=0.8]{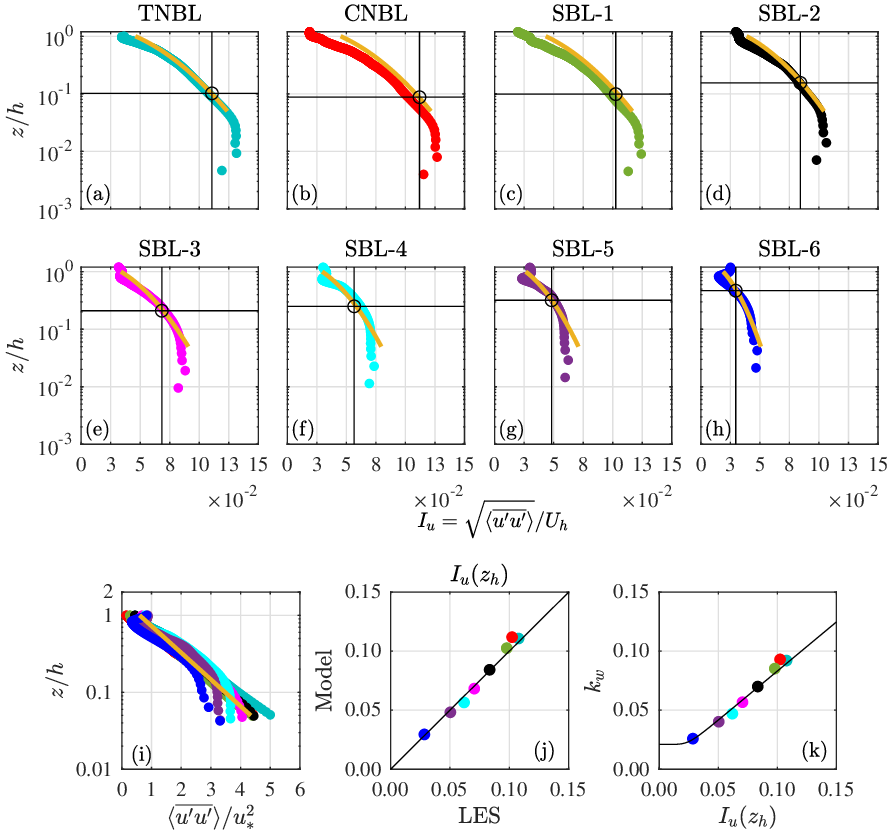}
    \caption{Plots depict semi-log vertical profiles of turbulence intensity $I_u=\sqrt{\mean{\overline{u^\prime u^\prime}}}/U_h$ predicted from Eq. \eqref{Iu_model} ($\protect\Iumodel$) and comparisons with LES data for (a) TNBL ($\protect\TNBLcirc$), (b) CNBL ($\protect\CNBLcirc$), (c) SBL-1 ($\protect\SBLacirc$), (d) SBL-2 ($\protect\SBLbcirc$), (e) SBL-3 ($\protect\SBLccirc$), (f) SBL-4 ($\protect\SBLdcirc$), (g) SBL-5 ($\protect\SBLecirc$), (h) SBL-6 ($\protect\SBLfcirc$) cases. Plot (i) shows normalized streamwise variance $\mean{\overline{u^\prime u^\prime}}/u_*^2$ from Eq. \eqref{upup_var} ($\protect\Iumodel$) compared with LES results following same color scheme as in plots (a)-(h) for the different ABL conditions. Plot (j) shows the variation of $I_u$ at the turbine hub height $(z_h)$ (marked by circle markers in each plot a-h) across the different ABL scenarios and its good agreement with LES results. Panel (k) shows the variation of $k_w$ as a function of $I_u(z_h)$ evaluated from Eq. \eqref{kw_CNBL_SBL_model}. The values of $k_w$ evaluated using the same equation for each ABL condition are shown in colored markers.} 
    \label{turb_intensity_fig}
\end{figure}

The value of $k_w$ depends on the turbulence intensity at hub height, $I_u(z_h)$, which is based on a logarithmic expression for the streamwise variance (Eq. \ref{upup_var}). As outlined in \S\ref{sec:wake_ABL_model}, the parameter $B_1$ is not expected to be universal and it depends on outer flow conditions.   Using the current LES data for various atmospheric conditions, shown in Figs. \ref{turb_intensity_fig}(a)-(h) for $I_u(z)$ and Fig. \ref{turb_intensity_fig}(i) for the normalized streamwise variance, $B_1$ is set to 0.6. Model predictions (yellow lines) for $I_u$ (Eq. \ref{Iu_model}) and $\mean{\overline{u^\prime u^\prime}}/u_*^2$ (Eq. \ref{upup_var}) are seen to be in good agreement with the LES results. It is noteworthy that the friction velocity ($u_*$) and the ABL flow velocity at hub height ($U_h$), used in the expression for $I_u$, are fully obtained from the analytical model discussed in \S\ref{sec:ABL} and not from LES, ensuring a self-contained determination of $I_u$.

The variation of $I_u(z_h)$ and $k_w$ across different atmospheric conditions is depicted in Figs. \ref{turb_intensity_fig}(j) \& (k), respectively. While the magnitude of $I_u(z_h)$ is comparable for both TNBL and CNBL scenarios, turbulence intensity decreases as the surface cooling rate increases. Consequently, Eq. \eqref{kw_CNBL_SBL_model} effectively captures this behavior by predicting a higher value of $k_w$ under neutral conditions, which decreases under strongly stable stratification. This stability-dependent $k_w$ is then used to calculate the wake width from Eq. \eqref{sigma_width} and the velocity deficit magnitude, $C(x)$, from Eq. \eqref{Cx}.

\subsection{Velocity deficit magnitude decay}\label{sec:Cx}
\begin{figure}[H]
    \centering
\includegraphics[scale=1]{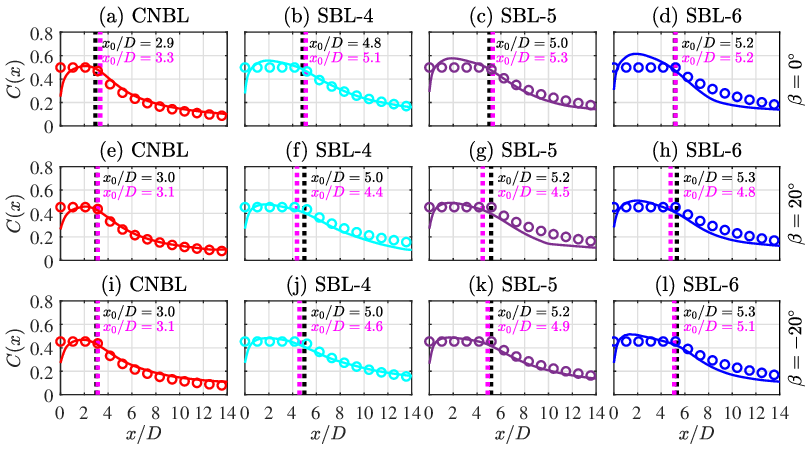}
    \caption{Plots depict decay of maximum velocity deficit $C(x)=\Delta u_{\text{max}}/U_h$ of wakes formed behind a wind turbine with yaw angles $\beta=0^\circ$ (a-d), $\beta=20^\circ$ (e-h) and $\beta=-20^\circ$ (i-l) placed under CNBL (a,e,i), SBL-4 (b,f,j), SBL-5 (c,g,k) and SBL-6 (d,h,l) atmospheric conditions. The predictions of $C(x)$ evaluated from Eq. \eqref{Cx} (circle markers) are compared to LES results (solid lines). The dotted black line represents the transition location $x_0/D$ evaluated from Eq. \eqref{x0_transition}.   The dotted magenta line is the transition location evaluated from LES. The precise values of $x_0/D$ evaluated from Eq. \eqref{x0_transition} and LES are shown as black and magenta colored text, respectively adjacent to the dotted lines.}
    \label{Cx_figure}
\end{figure}
Fig. \ref{Cx_figure} illustrates the downstream evolution of the maximum velocity deficit $C(x)$, for wind turbines with yaw angles $\beta = 0^\circ, \pm 20^\circ$ in CNBL and SBL-4, 5, and 6 atmospheric conditions.  $C(x)$, calculated using Eq. \eqref{Cx}, shows good agreement with LES results across all yaw angles, with only slight deviations near the rotor. 
In these plots, the black dotted vertical lines are the transition point, $x_0$, predicted by Eq. \eqref{x0_transition}.  The transition location derived from LES is marked by magenta dotted vertical lines. The $x_0/D$ from LES is determined as the position where 
$C(x)$ reaches approximately 95\% of the theoretical maximum velocity deficit magnitude ($2a$) in the potential core region. From these vertical dotted lines, we note that for CNBL conditions, the potential core extends about three diameters downstream of the wind turbine, whereas in SBL flows, it stretches up to five diameters. The model prediction of $x_0$ using Eq. \eqref{x0_transition} is in good agreement with LES results for the CNBL and SBL atmospheric conditions  although minor deviations ranging between $0.3D$ and $0.6D$ can be observed for some of the cases.

The observed agreement between the model and LES for $C(x)$ further validates the effectiveness of the model for the wake expansion rate, $k_w$, provided by Eq. \eqref{kw_CNBL_SBL_model}. The faster wake recovery in neutral conditions, characterized by a shorter $x_0$, is attributed to the more effective turbulent transport of momentum from the free-stream into the wake. In neutral ABL flows, where $I_u$ is higher, this momentum transport is more efficient, resulting in faster wake recovery. In contrast, SBL flows exhibit reduced $I_u$ due to stable stratification, leading to longer wake structures. In summary, using Eqs. \eqref{Cx} and \eqref{kw_CNBL_SBL_model}, we successfully capture the wake decay and expansion behavior for wind turbines in both CNBL and SBL conditions.

\subsection{Comparing entire velocity distributions from analytical model and LES}\label{sec:wake_model_vs_LES}

This section assesses the accuracy of model predictions from Eq. \eqref{du_model} for both yawed and unyawed wind turbines under CNBL and SBL-4, 5, and 6 atmospheric conditions by comparing them with LES results. Predictions for the unyawed turbine are detailed in \S\ref{yaw0_sec}, while those for yawed turbines are covered in \S\ref{yawpm20_sec}.

\subsubsection{Computing wake structures of unyawed ($\beta=0^\circ$) wind turbine in ABL flows}\label{yaw0_sec}

\begin{figure}[H]
    \centering
    \includegraphics[scale=1]{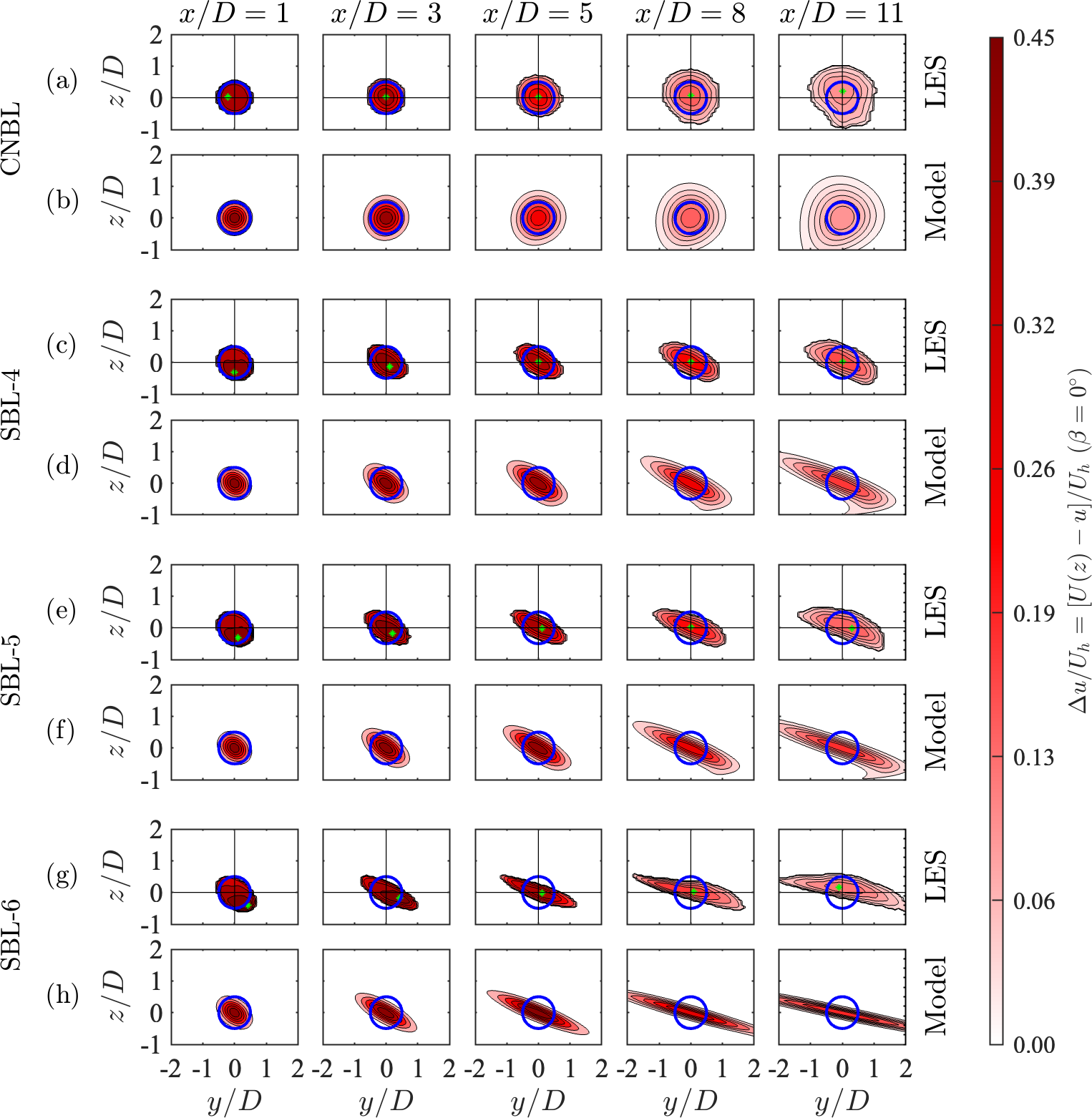}
    \caption{Comparison of $\Delta u/U_h$ predicted by Eq. \eqref{du_model} (b, d, f, h) with contours from LES (a, c, e, g) of an unyawed turbine placed in CNBL (a,b), SBL-4 (c,d), SBL-5 (e,f) and SBL-6 (g,h) atmospheric flows. 
    In the LES contours, the green marker indicates the location of the maximum velocity deficit, while the blue circle represents the wind turbine's rotor edge, with a diameter of $D=100$ m.}
    \label{fig:yaw0_du_LES_vs_model}
\end{figure}

The wake model predictions of the velocity deficit structure for an unyawed wind turbine under CNBL, SBL-4,5,6 atmospheric conditions are compared with LES results in Fig. \ref{fig:yaw0_du_LES_vs_model} at different downstream locations $x/D = [1, 3, 5, 8, 11]$. The LES contours of $\Delta u/U_h$ are shown in Figs. \ref{fig:yaw0_du_LES_vs_model}(a, c, e, g), and the model predictions are presented in Figs. \ref{fig:yaw0_du_LES_vs_model}(b, d, f, h).

As discussed in \S\ref{wake_LES}, these contours reveal that in the absence of yaw, wake deflection is driven solely by wind veer. As a result, veer-induced wake deflection is more pronounced in SBL flows than in CNBL flows, which is reflected in the LES contours (Figs. \ref{fig:yaw0_du_LES_vs_model}: a, c, e, g). The wake model predictions in Figs. \ref{fig:yaw0_du_LES_vs_model}(b, d, f, h) also capture these characteristics, producing symmetric wake shapes under neutral conditions and tilted wakes in SBL flows. Fig. \ref{fig:yaw0_du_LES_vs_model} also indicates that the model successfully replicates the wake decay behavior observed in the LES results.

Although the model effectively captures key features of the wake structure and changes due to veer and atmospheric effects, minor differences in the wake shape are observed compared to LES. For instance, the model assumes the peak velocity deficit always occurs at the rotor center, whereas the green markers from the LES results show that this point can deviate from the wake center. This change in the peak location can be attributed to cross-stream advection caused by secondary flows \citep{Narasimhan_et_al_JPCS_2024}. These secondary flow effects, visible in the LES contours, are not accounted for in the wake model. Nevertheless, the model reproduces main features of the wake distribution across a range of atmospheric conditions in a self-consistent and fully-predictive manner.

\begin{figure}[H]
    \centering
    \includegraphics[scale=0.8]{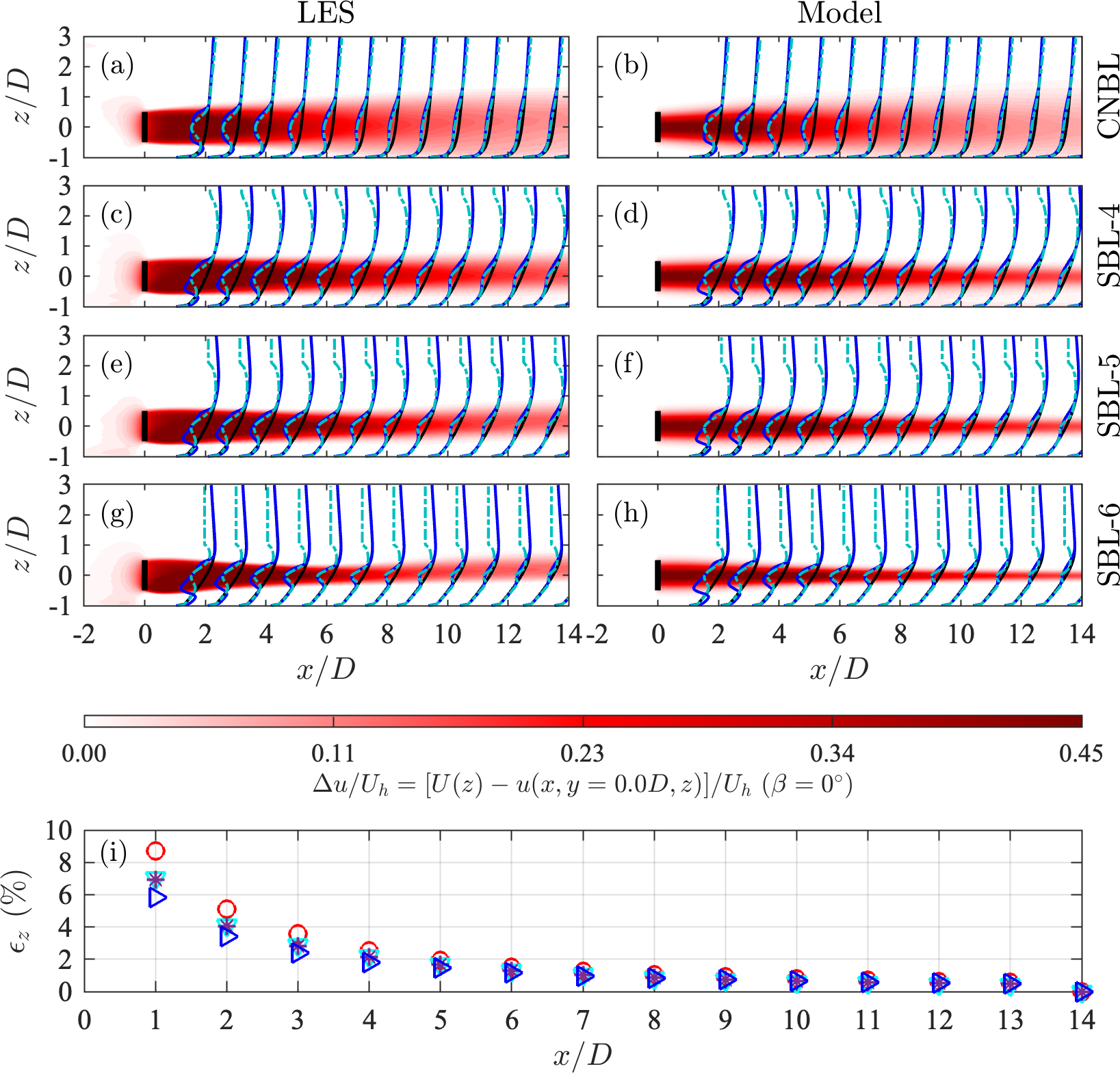}
    \caption{Contours of $\Delta u(x,y=0,z)/U_h$ evaluated from wake model (b,d,f,h) compared with LES (a,c,e,g) for unyawed turbine placed in CNBL (a,b), SBL-4 (c,d), SBL-5 (e,f), SBL-6 (g,h) ABL flows. The contour plots are overlaid with line plots of ABL inflow $U(z)$ ($\protect\blackline$) from LES, wake velocity profile $u(x,y/D=0,z)$ ($\protect\blueline$) from LES, wake velocity profile $u(x,y/D=0,z)$ ($\protect\cyandashdotline$) evaluated from the wake model. Figure (i) shows the mean absolute percentage error ($\epsilon_z$ from Eq. \ref{eps_z}) between the LES and model prediction of vertical profiles of $u(x,y,z)$ at downstream locations $x/D=[1-14]$  for CNBL $(\protect\CNBLcircopen)$, SBL-4 $(\protect\SBLdtriangle)$, SBL-5 $({\color{color6}*})$, and SBL-6 $(\protect\SBLftriangle)$ flows. }
    \label{fig:yaw0_du_LES_vs_model_xz}
\end{figure}

\begin{figure}[H]
    \centering
    \includegraphics[scale=0.8]{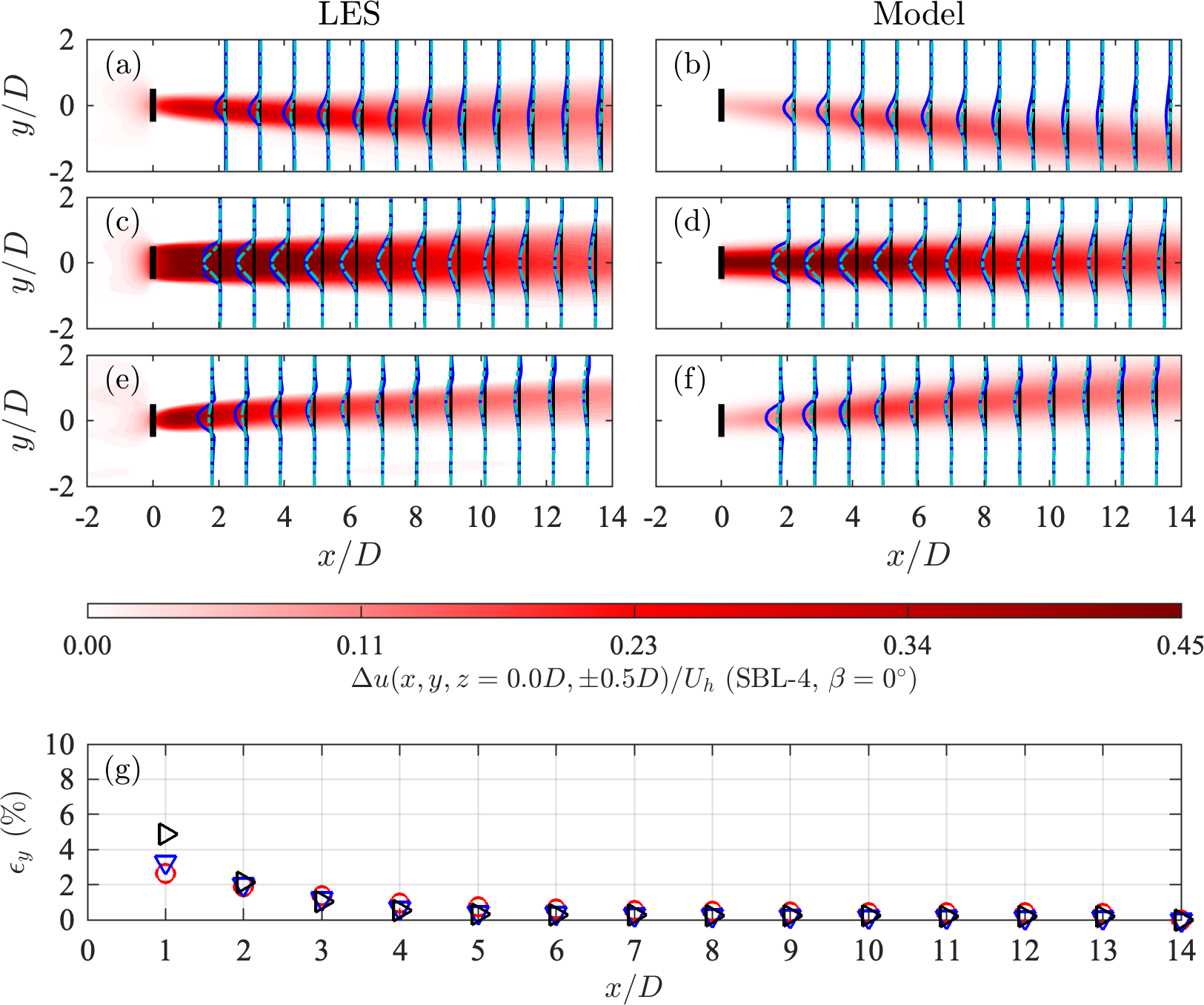}
    \caption{Contours of velocity deficit $\Delta u/U_h$ at vertical heights $z/D=0.5$ (a,b), $z/D=0$ (c,d) and $z/D=-0.5$ (e,f) evaluated from wake model (b,d,f) compared with LES (a,c,e) for unyawed turbine placed in SBL-4 atmospheric flow. The line plots on top of the contours show spanwise profiles of undisturbed ABL inflow velocity ($\protect\blackline$), flow velocity within wake region $(u(x,y,z)=U(z)-\Delta u(x,y,z))$ obtained from LES ($\protect\blueline$), and wake model ($\protect\cyandashdotline$). Figure (g) shows the mean absolute percentage error ($\epsilon_y$ from Eq. \ref{eps_y}) between the LES and model predictions of the spanwise profiles of $u(x,y,z)$ within the wake region plotted at vertical heights $z/D=-0.5$ ($\protect\trianglemarkereast$), $z/D=0$ ($\protect\bluetrianglemarker$), $z/D=0.5$ ($\protect\redcirc$)  and at downstream locations $x/D=[1-14]$.  }
    \label{fig:yaw0_du_LES_vs_model_xy}
\end{figure}

Figs. \ref{fig:yaw0_du_LES_vs_model_xz} and \ref{fig:yaw0_du_LES_vs_model_xy} show $\Delta u/U_h$ contours overlaid with wind velocity profiles within the wake region along $y/D=0$ and $z/D=0,\pm 0.5$ planes, respectively plotted from the LES and wake model. When evaluating the wake model for $u(x,y,z)=U(z)-\Delta u(x,y,z)$, we use $U(z)$ from the coupled Ekman-surface layer ABL model (Eq. \ref{U_vel}), and $\Delta u(x,y,z)$ is derived from the wake model (Eq. \ref{du_model}). 
Fig. \ref{fig:yaw0_du_LES_vs_model_xz} demonstrates that the $\Delta u/U_h$ contours predicted by the wake model (Fig. \ref{fig:yaw0_du_LES_vs_model_xz}: b,d,f,h) exhibit similar features to those obtained from the LES (Fig. \ref{fig:yaw0_du_LES_vs_model_xz}: a,c,e,g) across CNBL and SBL atmospheric conditions. In neutral conditions, the wake has a thicker vertical structure at downstream distances versus the wake in stable conditions. The wake model effectively captures this thinning of the wake further downstream.  This narrowing of the wake under stable conditions along the $y/D=0$ plane can be attributed to both the lower wake expansion rate and the tilting of the wake structure, caused by strong wind veer shearing the wake around the turbine's center. 

Fig. \ref{fig:yaw0_du_LES_vs_model_xz} also indicates the vertical variation of the flow velocity within the wake region, which are plotted over the velocity deficit contours at different downstream locations $(x/D=[1-14])$. The black line ($\blackline$) represents the undisturbed mean inflow ahead of the turbine. The velocity within the wake region, $u(x, y, z) = U(z) - \Delta u(x, y, z)$, as calculated from LES, is shown by the blue line ($\blueline$),
while the model prediction is indicated by cyan dash-dotted lines ($\cyandashdotline$). The model's velocity prediction $u(x, y, z)$ is obtained by subtracting the predicted $\Delta u(x, y, z)$ (Eq. \ref{du_model}) from the inflow ABL velocity $U(z)$, obtained from the ABL wind model (Eq. \ref{U_vel}).

To assess the model’s accuracy, panel (i) in Fig. \ref{fig:yaw0_du_LES_vs_model_xz} and panel (g) in Fig. \ref{fig:yaw0_du_LES_vs_model_xy} displays the mean absolute percentage error (MAPE) between the LES and model predictions of the vertical and horizontal (spanwise) profiles of $u(x,y,z)$ at various $x/D$ locations, respectively. In Fig. \ref{fig:yaw0_du_LES_vs_model_xz}, the MAPE of the vertical profiles of velocity at a given streamwise location is defined as 
\begin{align}
\epsilon_z(x) = \frac{1}{N^\prime_z} \sum_{k=1}^{N^\prime_z} 
\left| \frac{u(x,y=0,z_k)_{\text{LES}} - u(x,y=0,z_k)_{\text{model}}}{u(x,y=0,z_k)_{\text{LES}}} \right| \times 100,\label{eps_z}
\end{align}
where the sum extends over all $N_z^\prime$ number of LES grid points between the ground and a height of $z=3D$. 
Similarly, in Fig. \ref{fig:yaw0_du_LES_vs_model_xy}, the error in the horizontal and spanwise profiles of velocity at a given streamwise location and $-2<y/D<2$ is measured by
\begin{align}
\epsilon_y(x) = \frac{1}{N^\prime_y} \sum_{j=1}^{N^\prime_y} 
\left| \frac{u(x,y_j,z/D=0,\pm 0.5)_{\text{LES}} - u(x,y_j,z/D=0,\pm 0.5)_{\text{model}}}{u(x,y_j,z=0,\pm 0.5)_{\text{LES}}} \right| \times 100.\label{eps_y}
\end{align}
In Eq. \eqref{eps_y}, the sum is over all LES grid points $j$ in the transverse direction such that $-2<y_j/D<2$. 
Fig. \ref{fig:yaw0_du_LES_vs_model_xz} (i) depicts $\epsilon_z(x)$ for CNBL, SBL-4,5, \& 6 atmospheric conditions using different colored markers.  The MAPE plot confirms good agreement between the model and LES at greater downstream distances where the errors are less than $2\%$ while the errors are above $5\%$ for regions within one diameter downstream of the wind turbine across all atmospheric conditions. This improved agreement further downstream is expected since the wake model assumes a negligible impact from the pressure gradient, a condition valid only farther from the near-wake region. Closer to the turbine, the velocity profile exhibits a top-hat shape, which becomes smoothed further downstream. Notably, the model captures the wake recovery seen in  LES results quite well. However, some differences are observed in the velocity structure between the model and LES, particularly in the SBL cases above the turbines. These discrepancies arise due to the zero-turbulent stress assumption made in the ABL model above the boundary layer height. In LES, residual stresses remain above the SBL height, potentially due to inertial oscillations, which are not included in the analytical model. Despite these differences, the extended analytical wake model effectively captures the veering low-level jet structure and its impact on the wind turbine wake.
In Fig. \ref{fig:yaw0_du_LES_vs_model_xy}, the top view of the velocity deficit contours at hub height ($z/D=0$), as well as $0.5D$ below ($z/D=-0.5$) and above ($z/D=0.5$) the hub height, are presented for a wind turbine in SBL-4 flow. The LES contours (Fig. \ref{fig:yaw0_du_LES_vs_model_xy}: a, c, e) show that the wake deflects in opposite directions above and below the hub height: along the positive spanwise direction below and the negative spanwise direction above, while there is no wake deflection at hub height. This deflection pattern arises from the wind veer, which is zero at the hub height but changes direction above and below. The line plots overlaid on the contours at various downstream locations follow the same convention as Fig. \ref{fig:yaw0_du_LES_vs_model_xz}.
 The MAPEs ($\epsilon_y$) between the LES ($\protect\blackline$) and model predictions ($\protect\cyandashdotline$) of the spanwise profiles of the flow velocity are shown in Fig. \ref{fig:yaw0_du_LES_vs_model_xy}(g). Again the model predictions of the velocity agree with the LES results at far downstream locations where the MAPEs are less than 2\% while the errors are around 5\% closer to the turbine. These discrepancies can again be attributed to  pressure gradient effects in the LES, which are assumed to be negligible in the wake model.

We may conclude that the proposed analytical wake model  predicts the wake structure with good accuracy for an unyawed wind turbine across conventionally-neutral and stable atmospheric conditions. In the following subsection, we will explore the model's ability to predict wake structures for yawed wind turbines placed in ABL flows.

\subsubsection{Computing wake structures behind yawed ($\beta=\pm 20^\circ$) wind turbine placed in ABL flows}\label{yawpm20_sec}

\begin{figure}[H]
    \centering
    \includegraphics[scale=1]{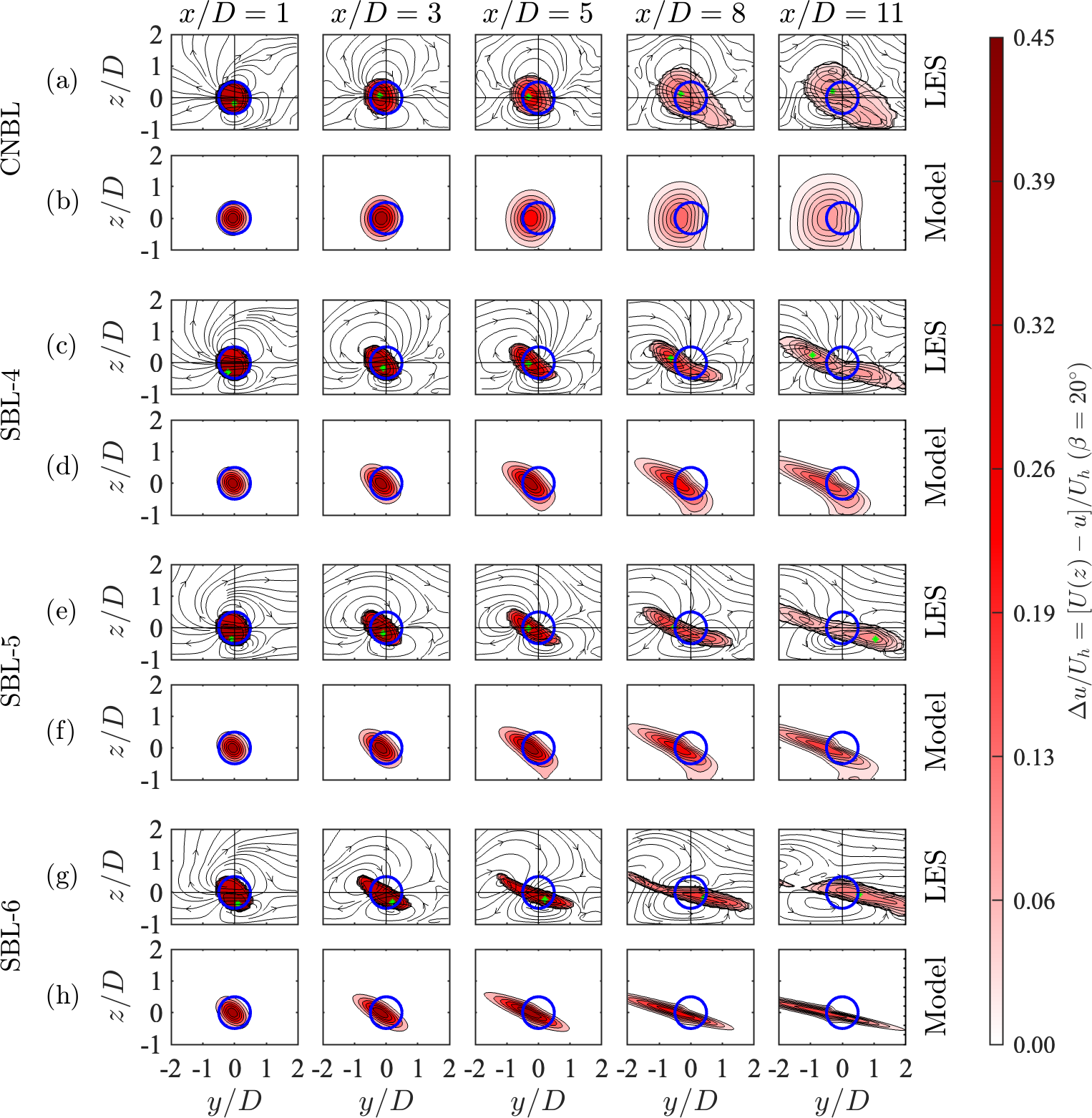}
    \caption{Comparison of $\Delta u/U_h$ predicted by Eq. \eqref{du_model} (b, d, f, h) with contours from LES (a, c, e, g) at locations $x/D=[1,3,5,8,11]$ for a $\beta=20^\circ$ yawed turbine placed in CNBL (a,b), SBL-4 (c,d), SBL-5 (e,f) and SBL-6 (g,h) atmospheric flows. The LES contours are overlaid with cross-stream ($v$-$w$) velocity streamlines. In the LES contours, the green marker indicates the location of the maximum velocity deficit, while the blue circle represents the wind turbine's rotor edge, with a diameter of $D=100$ m.}
    \label{fig:yaw20_du_LES_vs_model}
\end{figure}

\begin{figure}[H]
    \centering
    \includegraphics[scale=1]{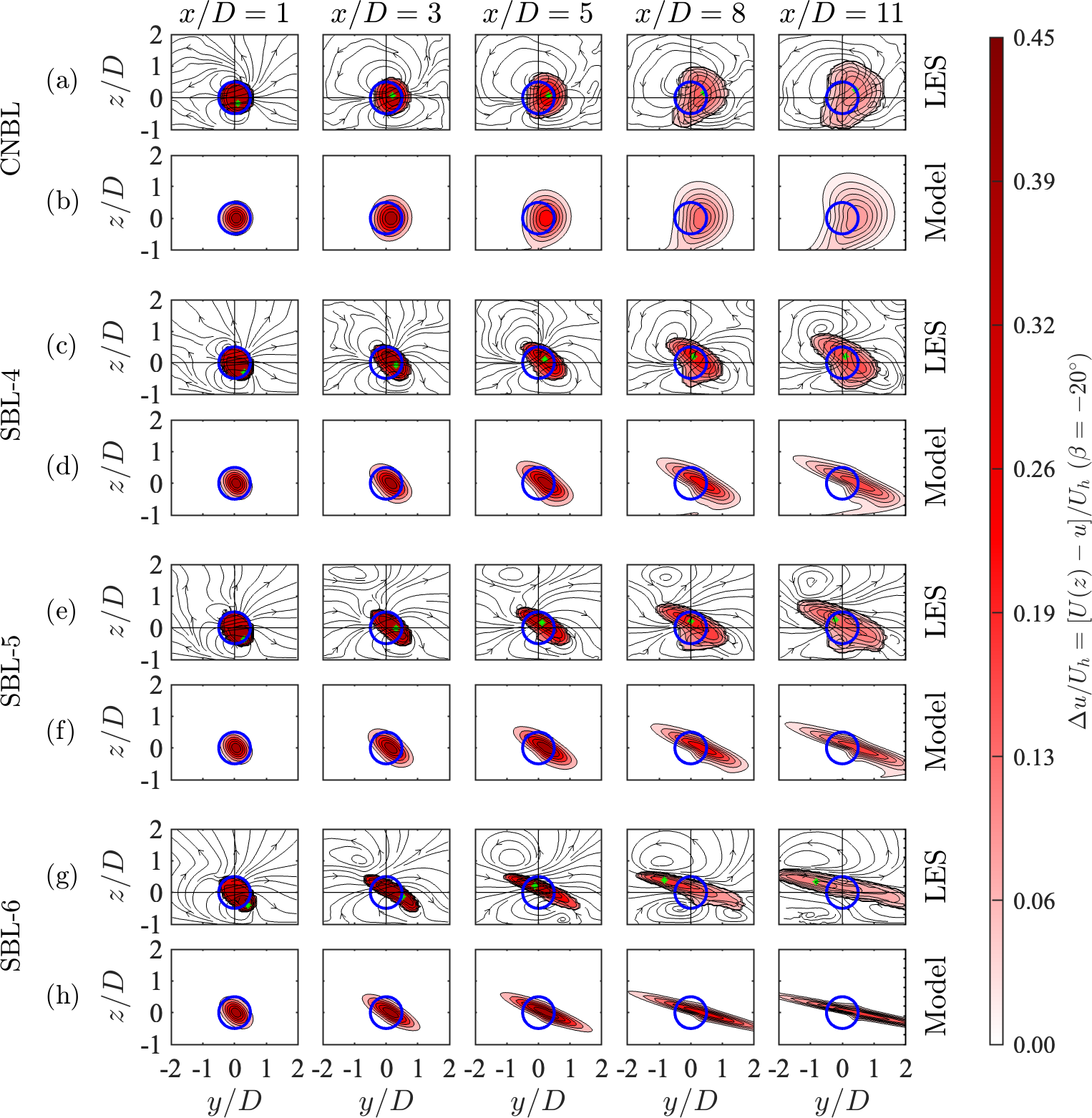}
    \caption{Comparison of $\Delta u/U_h$ predicted by Eq. \eqref{du_model} (b, d, f, h) with contours from LES (a, c, e, g) at locations $x/D=[1,3,5,8,11]$ for a $\beta=-20^\circ$ yawed turbine placed in CNBL (a,b), SBL-4 (c,d), SBL-5 (e,f) and SBL-6 (g,h) atmospheric flows. The LES contours are overlaid with cross-stream ($v$-$w$) velocity streamlines. In the LES contours, the green marker indicates the location of the maximum velocity deficit, while the blue circle represents the wind turbine's rotor edge, with a diameter of $D=100$ m.}
    \label{fig:yaw-20_du_LES_vs_model}
\end{figure}

In Figs. \ref{fig:yaw20_du_LES_vs_model} and \ref{fig:yaw-20_du_LES_vs_model}, we compare the wake structures obtained from the LES and the model at selected downstream locations for yawed ($\beta=\pm 20^\circ$) wind turbines placed in CNBL, SBL-4, 5, \& 6 ABL flows. In both figures, the velocity deficit contours from the LES are overlaid with $v$-$w$ cross-stream velocity streamlines for visualizing the secondary flow features. For better visualization of these secondary flows, the background veer velocity $V(z)$ has been subtracted from the spanwise flow velocity. 

The yaw-induced wake deflection is more pronounced for turbines operating in CNBL flow, as shown in Figs. \ref{fig:yaw20_du_LES_vs_model}  and  \ref{fig:yaw-20_du_LES_vs_model}: (a). The cross-stream velocity streamlines highlight the counter-rotating vortex pair (CVP) structure responsible for wake deflection under yawed conditions. The corresponding velocity deficit contours (Figs. \ref{fig:yaw20_du_LES_vs_model} and \ref{fig:yaw-20_du_LES_vs_model} (b)) from the wake model similarly capture the deflected and curled wake structure for yawed turbines in CNBL flow. However, minor discrepancies can be observed in the lower part for the $\beta=20^\circ$ yawed turbine, where the LES velocity deficit shows greater deflection toward the positive spanwise direction than the model prediction. This deviation is attributed to the misalignment in the spanwise locations of the top and bottom vortices of the CVP. For example, in Fig. \ref{fig:yaw20_du_LES_vs_model}(a) at $x/D=11$, the bottom vortex is advected more towards the left than the top vortex. This misalignment is caused by the lower streamwise advection velocity near the ground due to shear in the $U(z)$ ABL profile \citep{Narasimhan_et_al_2022}. The time a fluid particle has evolved after interacting with the turbine can be estimated as $t=x/U(z)$, where $x$ represents a given streamwise distance $x$ from the turbine. Therefore the bottom vortex has more time than the top vortex to advect along the spanwise direction induced than the top vortex \citep{Narasimhan_et_al_2022} leading to misalignment in the spanwise location of the vortices. Although the wind veer strength is minimal in the CNBL flow, the misaligned bottom vortex locally advects the wake towards the right causing a tilted wake structure. 
The wake model does not account for the effects of this vortex misalignment, so the spanwise-deflected and curled wake structure remains symmetric around the hub height in the model prediction. For the yawed turbine at $\beta=-20^\circ$  in the CNBL flow, the opposite happens where the bottom vortex (Fig. \ref{fig:yaw-20_du_LES_vs_model}(a) at $x/D=11$) is advected more towards the left than the top vortex. This behavior is again attributed to ABL shear and the tilted CVP structure determining the wake shape in this scenario. The model prediction of the velocity deficit contour for $\beta=-20^\circ$ in CNBL flow (Fig.  \ref{fig:yaw-20_du_LES_vs_model}(b)) closely resemble the LES contours.

For the SBL-4, 5, and 6 cases, the combined effects of wind veer and yaw-induced deflection significantly influence the wake shape. As shown in Figs. \ref{fig:yaw20_du_LES_vs_model} (c), (e), and (g), the wake behind the $\beta=20^\circ$ yawed turbine is sheared by the wind veer in each of the respective cases. In addition to the wake tilting, the CVP vortices are advected by the wind veer, with the sidewash velocity from the vortices directed downward, disrupting the wake and resulting in complex wake structures. Similarly, Fig. \ref{fig:yaw-20_du_LES_vs_model} (c), (e), and (g) present the LES results for the $\beta=-20^\circ$ yawed turbine in the SBL-4, 5, and 6 cases, respectively, where the wake shape is again influenced by the combined effects of yaw and wind veer. The CVP vortices are tilted, but in contrast to the $\beta=20^\circ$ case, the sidewash velocity is directed upward, away from the ground. Although the wake model does not account for the tilting of the vortices by wind veer, it still reasonably captures the complex wake structures under the combined influence of yaw and wind veer for both $\beta=\pm 20^\circ$ yawed turbines across different ABL flows with varying wind veer strengths as depicted in Figs. \ref{fig:yaw20_du_LES_vs_model} \& \ref{fig:yaw-20_du_LES_vs_model} (d), (f), (h).

\begin{figure}[H]
    \centering
    \includegraphics[scale=0.8]{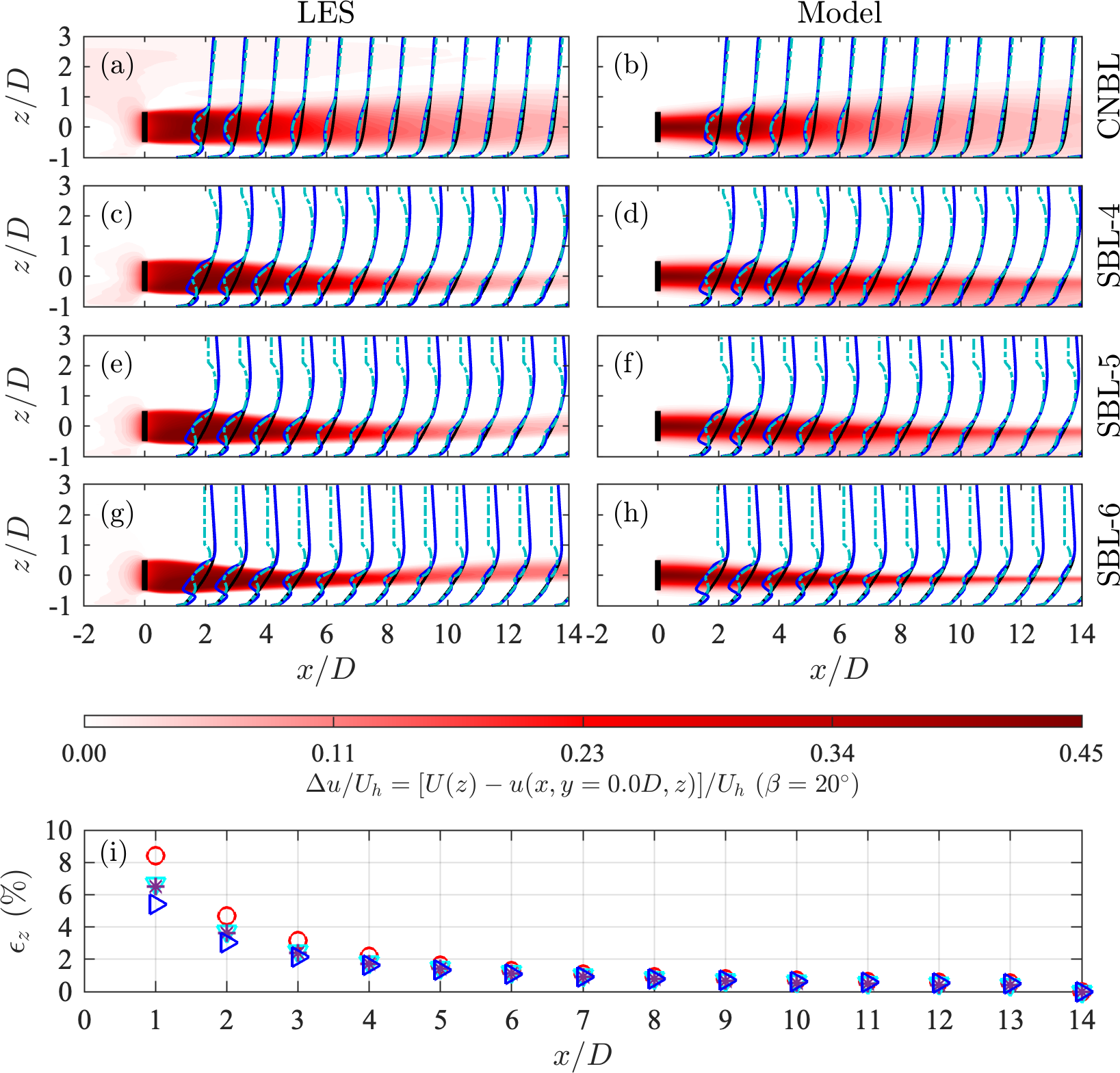}
    \caption{Contours of $\Delta u(x,y=0,z)/U_h$ evaluated from wake model (b,d,f,h) compared with LES (a,c,e,g) for $\beta=20^\circ$ yawed turbine placed in CNBL (a,b), SBL-4 (c,d), SBL-5 (e,f), SBL-6 (g,h) ABL flows. The contour plots are overlaid with line plots of ABL inflow $U(z)$ ($\protect\blackline$) from LES, wake velocity profile $u(x,y/D=0,z)$ ($\protect\blueline$) from LES, wake velocity profile $u(x,y/D=0,z)$ ($\protect\cyandashdotline$) evaluated from the wake model. Figure (i) shows the mean absolute percentage error ($\epsilon_z$ from Eq. \ref{eps_z}) between the LES and model prediction of vertical profiles of $u(x,y,z)$ at downstream locations $x/D=[1-14]$  for CNBL $(\protect\CNBLcircopen)$, SBL-4 $(\protect\SBLdtriangle)$, SBL-5 $({\color{color6}*})$, and SBL-6 $(\protect\SBLftriangle)$ flows.}
    \label{fig:yaw20_du_LES_vs_model_xz}
\end{figure}

\begin{figure}[H]
    \centering
    \includegraphics[scale=0.8]{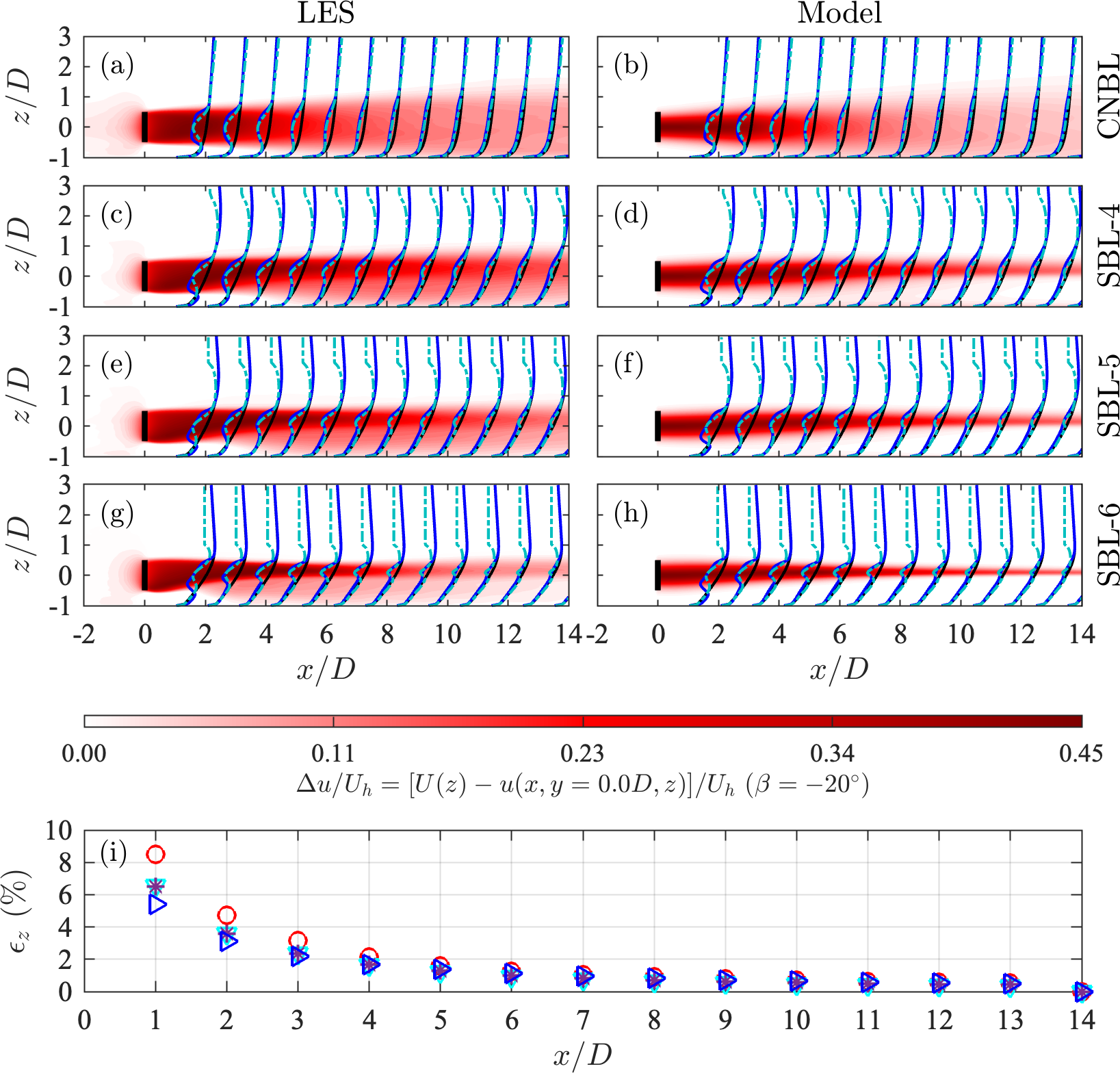}
    \caption{Contours of $\Delta u(x,y=0,z)/U_h$ evaluated from wake model (b,d,f,h) compared with LES (a,c,e,g) for $\beta=-20^\circ$ yawed turbine placed in CNBL (a,b), SBL-4 (c,d), SBL-5 (e,f), SBL-6 (g,h) ABL flows. The contour plots are overlaid with line plots of ABL inflow $U(z)$ ($\protect\blackline$) from LES, wake velocity profile $u(x,y/D=0,z)$ ($\protect\blueline$) from LES, wake velocity profile $u(x,y/D=0,z)$ ($\protect\cyandashdotline$) evaluated from the wake model. Figure (i) shows the mean absolute percentage error ($\epsilon_z$ from Eq. \ref{eps_z}) between the LES and model prediction of vertical profiles of $u(x,y,z)$ at downstream locations $x/D=[1-14]$  for CNBL $(\protect\CNBLcircopen)$, SBL-4 $(\protect\SBLdtriangle)$, SBL-5 $({\color{color6}*})$, and SBL-6 $(\protect\SBLftriangle)$ flows.}
    \label{fig:yaw-20_du_LES_vs_model_xz}
\end{figure}

\begin{figure}[H]
    \centering
    \includegraphics[scale=0.8]{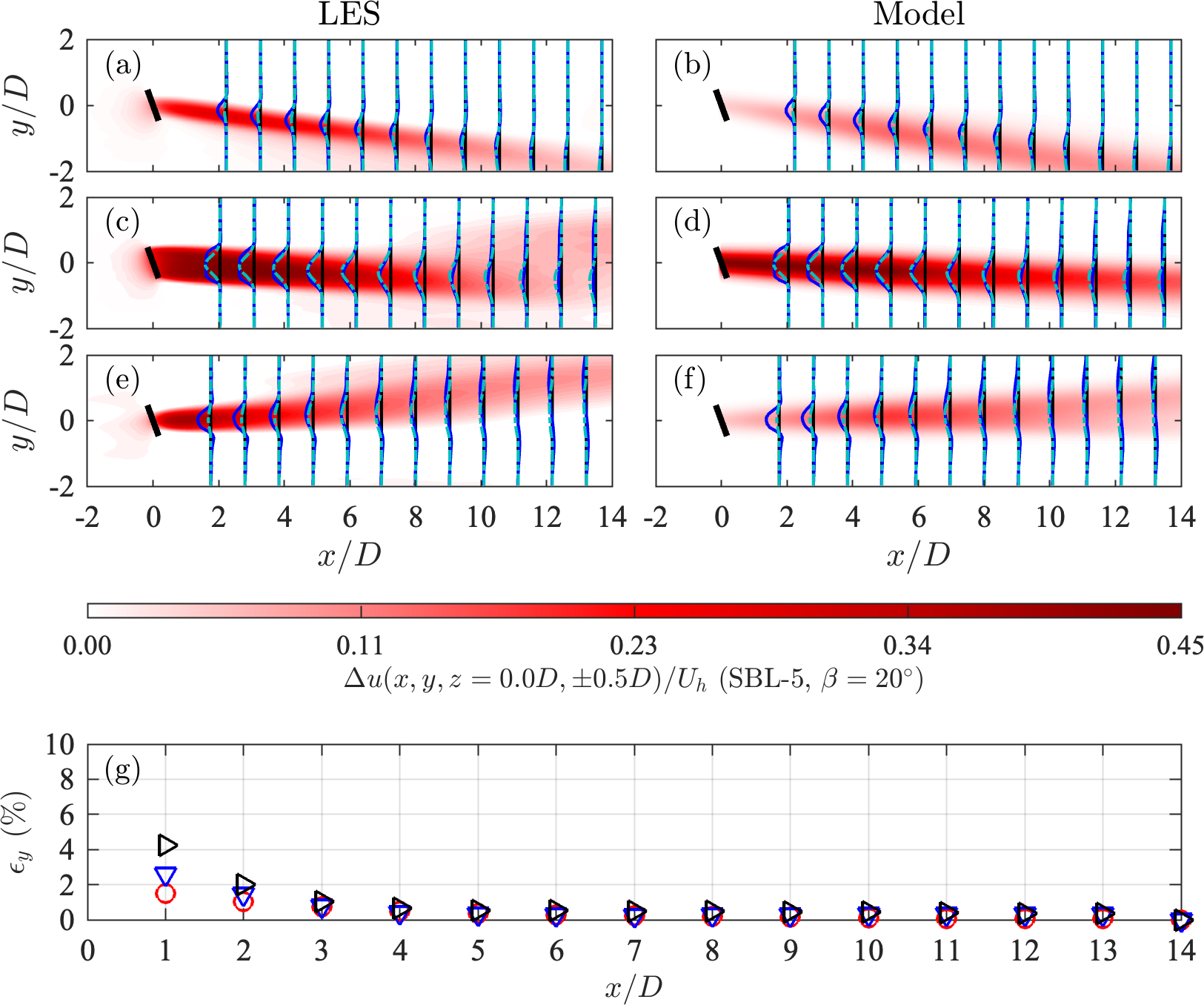}
        \caption{Contours of velocity deficit $\Delta u/U_h$ at vertical heights $z/D=0.5$ (a,b), $z/D=0$ (c,d) and $z/D=-0.5$ (e,f) evaluated from wake model (b,d,f) compared with LES (a,c,e) for $\beta=20^\circ$ yawed turbine placed in SBL-5 atmospheric flow. The line plots on top of the contours show spanwise profiles of undisturbed ABL inflow velocity ($\protect\blackline$), flow velocity within wake region $(u(x,y,z)=U(z)-\Delta u(x,y,z))$ obtained from LES ($\protect\blueline$) , and wake model ($\protect\cyandashdotline$). Figure (g) shows the mean absolute percentage error ($\epsilon_y$ from Eq. \ref{eps_y}) between the LES and model predictions of the spanwise profiles of $u(x,y,z)$ within the wake region plotted at vertical heights $z/D=-0.5$ ($\protect\trianglemarkereast$), $z/D=0$ ($\protect\bluetrianglemarker$), $z/D=0.5$ ($\protect\redcirc$)  and at downstream locations $x/D=[1-14]$.  }
    \label{fig:yaw20_du_LES_vs_model_xy}
\end{figure}

\begin{figure}[H]
    \centering
    \includegraphics[scale=0.8]{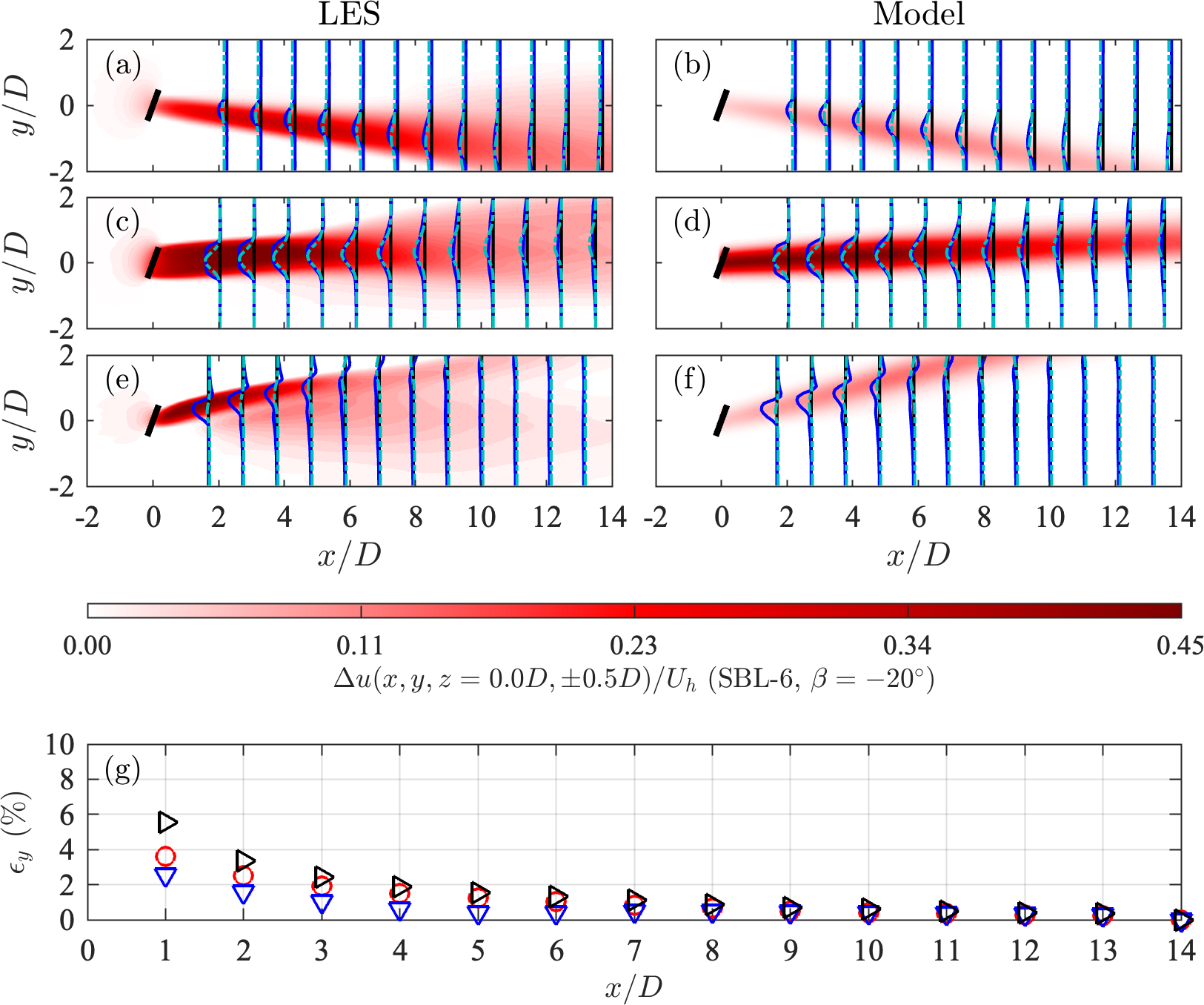}
    \caption{Contours of velocity deficit $\Delta u/U_h$ at vertical heights $z/D=0.5$ (a,b), $z/D=0$ (c,d) and $z/D=-0.5$ (e,f) evaluated from wake model (b,d,f) compared with LES (a,c,e) for $\beta=-20^\circ$ yawed turbine placed in SBL-6 atmospheric flow. The line plots on top of the contours show spanwise profiles of undisturbed ABL inflow velocity ($\protect\blackline$), flow velocity within wake region $(u(x,y,z)=U(z)-\Delta u(x,y,z))$ obtained from LES ($\protect\blueline$), and wake model ($\protect\cyandashdotline$). Figure (g) shows the mean absolute percentage error ($\epsilon_y$ from Eq. \ref{eps_y}) between the LES and model predictions of the spanwise profiles of $u(x,y,z)$ within the wake region plotted at vertical heights $z/D=-0.5$ ($\protect\trianglemarkereast$), $z/D=0$ ($\protect\bluetrianglemarker$), $z/D=0.5$ ($\protect\redcirc$)  and at downstream locations $x/D=[1-14]$.  }
    \label{fig:yaw-20_du_LES_vs_model_xy}
\end{figure}

Figs. \ref{fig:yaw20_du_LES_vs_model_xz} and \ref{fig:yaw-20_du_LES_vs_model_xz}  present a quantitative comparison between the LES results and the corresponding wake model predictions of the velocity deficit at the $y/D=0$ plane for turbines yawed at $\beta=20^\circ$ and $\beta=-20^\circ$, respectively. In each figure, panels (a), (c), (e), and (g) show the LES velocity deficit contours for yawed wind turbines operating in the CNBL and SBL-4, 5, and 6 conditions, respectively. 
The corresponding wake model predictions under these different ABL scenarios are shown in panels (b), (d), (f), and (h). The velocity profiles are overlaid as line plots on the wake contours in both figures.  In panel (i) of both Figs. \ref{fig:yaw20_du_LES_vs_model_xz} and \ref{fig:yaw-20_du_LES_vs_model_xz}, the MAPE ($\epsilon_z$ from Eq. \ref{eps_z}) decreases gradually, starting from 9\% just behind the wind turbine and dropping to below 2\% at downstream distances where $x/D > 5$. This highlights the ability of the wake model to correctly capture the far-wake behavior at greater distances downstream of yawed wind turbines across different atmospheric conditions. As with the unyawed case, the discrepancies between the model predictions and LES results near the yawed wind turbines can be attributed to the assumption that pressure gradient effects are negligible—a condition that is only valid at large downstream distances from the turbine.
Figure\ref{fig:yaw20_du_LES_vs_model_xy} compares the model predictions with LES at hub height, as well as at locations $0.5D$ below and above the hub height
for a $\beta=20^\circ$ yawed wind turbine in SBL-5 flow. Fig. \ref{fig:yaw-20_du_LES_vs_model_xy} shows the same results for a turbine yawed at  $\beta=-20^\circ$ in SBL-6 flow.  In both figures the spanwise profiles of the flow velocity in the wake region are overlaid on the contours, allowing for direct comparison between the wake model and LES results. In each figure, LES contours are shown in panels (a), (c), and (e), while the corresponding wake model contours are plotted in panels (b), (d), and (f). Panel (g) in each figure presents the MAPE ($\epsilon_y$ from Eq. \ref{eps_y}) between the wake model predictions and LES results for the flow velocity profiles at various downstream locations behind the yawed turbines.
Since the wind veer velocity is zero at hub height, the wake deflection is in the negative spanwise direction for a $\beta=20^\circ$ yawed turbine and in the positive spanwise direction for a $\beta=-20^\circ$ yawed turbine as shown in Figs. \ref{fig:yaw20_du_LES_vs_model_xy} (c) and \ref{fig:yaw-20_du_LES_vs_model_xy} (c), respectively.
In contrast, the wake deflection below and above hub height is driven by the wind veer velocity of the respective ABL flows. The model predictions are in good agreement with the LES results, as confirmed by the MAPE values of around 2-6\% just behind the turbine, which decrease to less than 2\% further downstream. 

Results shown in this section demonstrate the capability and versatility of the proposed analytical wake model in predicting various wake distributions and shapes for wind turbines subjected to both yaw and wind veer effects under conventionally neutral and stable atmospheric conditions.

\subsection{Power losses due to wake interactions: comparing LES and model predictions}
 \label{sec:power_wake_loss}

In this section, we evaluate the effectiveness of the proposed analytical wake model in predicting power loss due to wake interactions between wind turbines. We consider a scenario where a downstream (unyawed) wind turbine is positioned in the wake of an upstream turbine exposed to an undisturbed ABL inflow. The upstream turbine is yawed at angles of $\beta = 0^\circ, \pm 20^\circ$ and is placed within CNBL, SBL-4, 5, 6 atmospheric flows. We then compute the normalized power output, $P(x_T, y_T)/P_0$,
of the downstream turbine as a function of its position (streamwise and spanwise coordinates ($x_T,y_T$) relative to the upstream turbine, where $P_0$ is the reference power output of an upstream (unyawed) turbine subjected to an undisturbed ABL flow.

The reference power output $P_0$ for the unyawed wind turbine is assumed to be given by
\begin{align}
P_0&=\frac{1}{2}\rho \pi R^2 C_T^\prime [U_h(1-a)]^3 \label{P0_ref}.
\end{align}

The power output of a downstream wind turbine located at an arbitrary position $(x_T, y_T)$ that overlaps with  
the wake of an upstream turbine is defined as 
\begin{align}
P(x_T, y_T)&=\frac{1}{2}\rho \pi R^2 C_T^\prime U_d(x_T,y_T)^3 \label{P_hyp}.
\end{align}
In this equation, $U_d(x_T, y_T)$ represents the disk-averaged velocity at the specified $(x_T, y_T)$ location and is calculated as 
\begin{align}
U_d(x_T,y_T)&=(1-a)\frac{1}{\pi R^2}\int_y\int_z u(x_T,y-y_T, z) \ dz \ dy.\label{Ud}
\end{align}
Equation \eqref{Ud} involves averaging the velocity $u(x_T, y - y_T, z)$ over the rotor area. Here, the flow velocity is expressed as $u(x_T, y - y_T, z) = U(z) - \Delta u(x_T, y - y_T, z)$, where $U(z)$ is determined from the coupled Ekman-surface layer ABL  model described in \S\ref{sec:ABL}, and $\Delta u(x_T, y - y_T, z)$ is obtained from the wake model. The power ratio $P(x_T, y_T)/P_0$ can be written as
\begin{align}
\frac{P(x_T,y_T)}{P_0}=\frac{1}{U_h^3}\left[\frac{1}{\pi R^2}\int_y\int_z \left[U(z)-\Delta u(x_T,y-y_T,z)\right]  \ dy \ dz \right]^3.\label{P_ratio}
\end{align}

To demonstrate the improved predictive capability of the proposed new wake model (referred to as M-1) for power estimates, we compare its power predictions with those from two alternative models: the \textcite{bastankhah_et_al_2022}'s vortex sheet-based wake model without veer correction (referred to as M-2), and the widely used Gaussian wake model~\citep{bastankhah2014} (referred to as M-3), in its original unmodified form which does not account for yaw or veer effects. We first compare the predicted and LES streamwise mean velocity distributions at various downstream locations in cross-stream planes whose rotor-area integration yields the power predictions. Secondly, we analyze power predictions.   

\subsubsection{Streamwise velocity in cross-flow planes: model predictions and comparisons with LES}

In this section, we compare the contours of the velocity field $u(x, y, z)$ behind a wind turbine predicted by the M-1, M-2, and M-3 models and compare them with corresponding LES results, for three-yaw scenarios across CNBL and SBL atmospheric conditions.

\begin{figure}[H]
    \centering
    \includegraphics[scale=1]{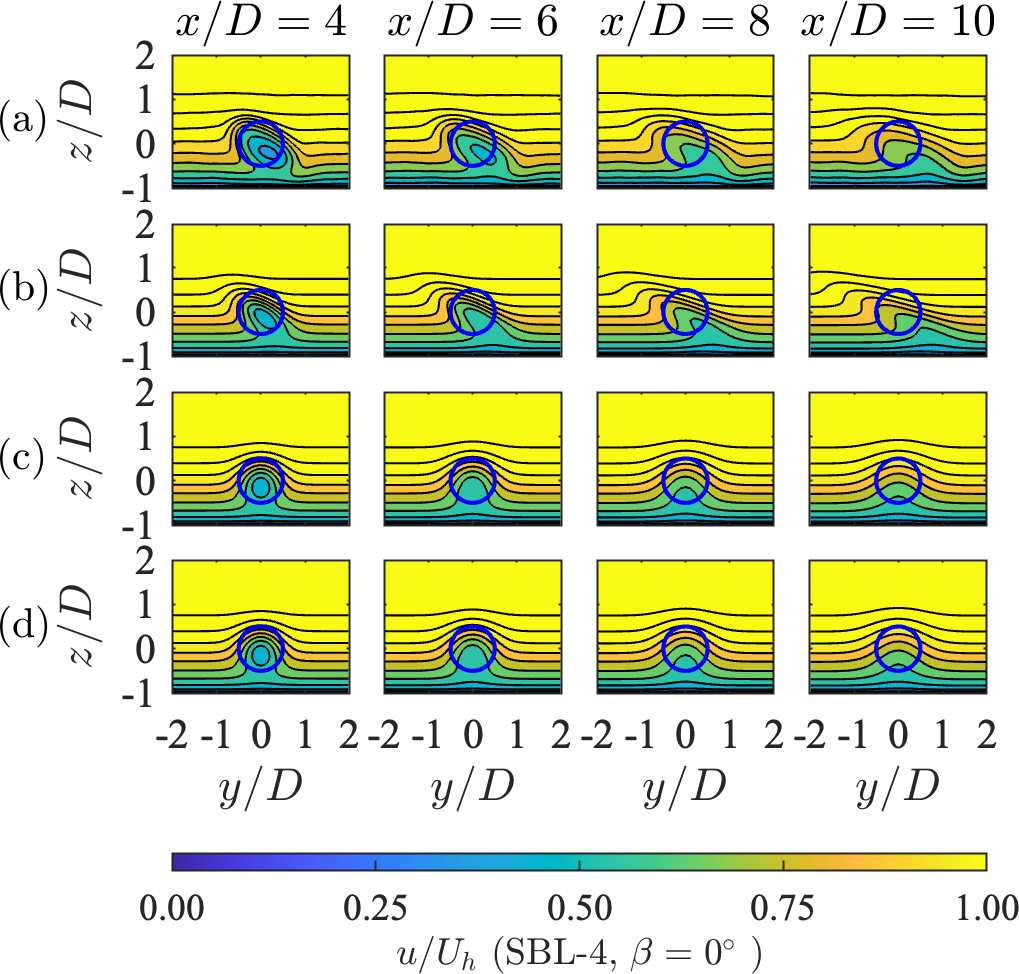}
    \caption{Contour plots of streamwise velocity  behind an unyawed wind turbine in SBL-4 flow at streamwise locations $x/D=[4,6,8,10]$ from (a) LES, (b) the veer-corrected vortex sheet-based wake model (M-1), (c) the vortex sheet-based wake model without veer correction (M-2), and (d) the Gaussian wake model (M-3).} \label{u_yaw0_SBL4_M1M2M3}
\end{figure}

\begin{figure}[H]
    \centering
    \includegraphics[scale=1]{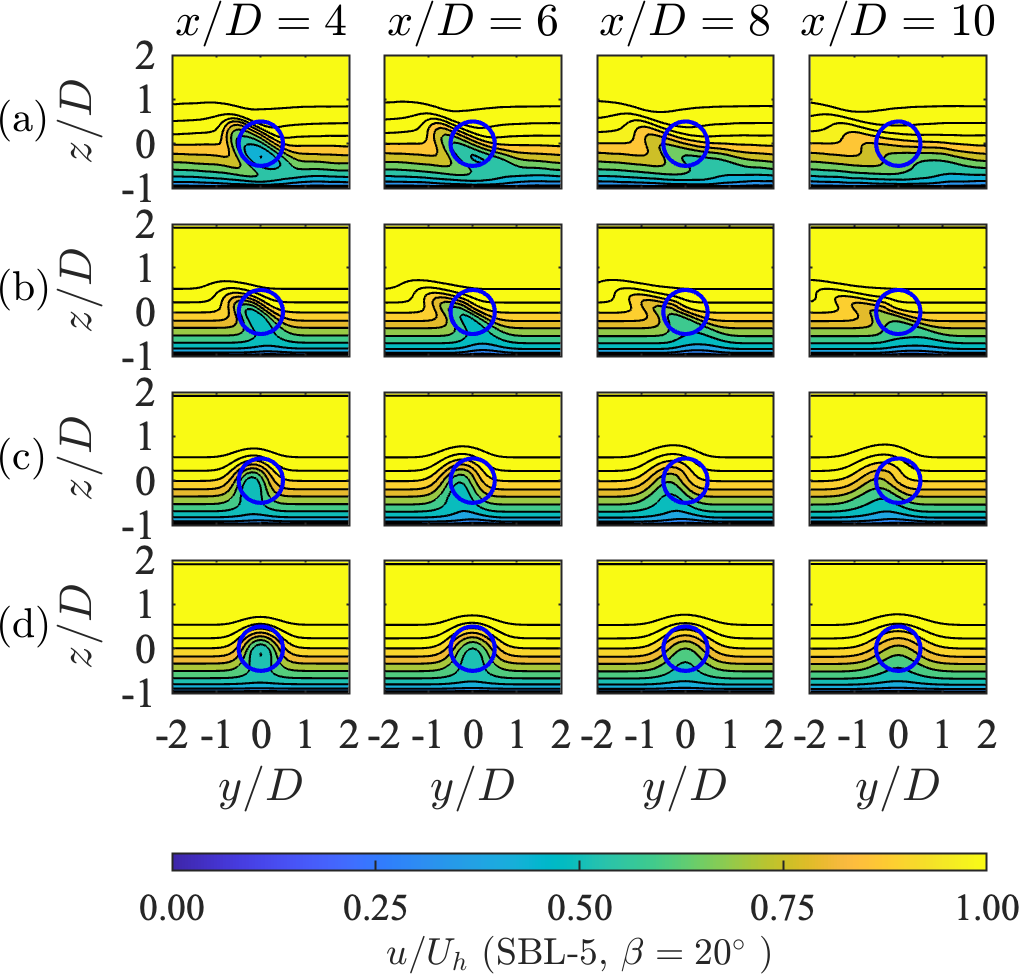}
    \caption{Contour plots of velocity field behind a  $\beta=20^\circ$ yawed wind turbine in SBL-5 flow at streamwise locations $x/D=[4,6,8,10]$ from (a) LES, (b) veer-corrected vortex sheet-based wake model (M-1), (c) the vortex sheet-based wake model without veer correction (M-2), and (d) the Gaussian wake model (M-3).} \label{u_yaw20_SBL5_M1M2M3}
\end{figure}

\begin{figure}[H]
    \centering
    \includegraphics[scale=1]{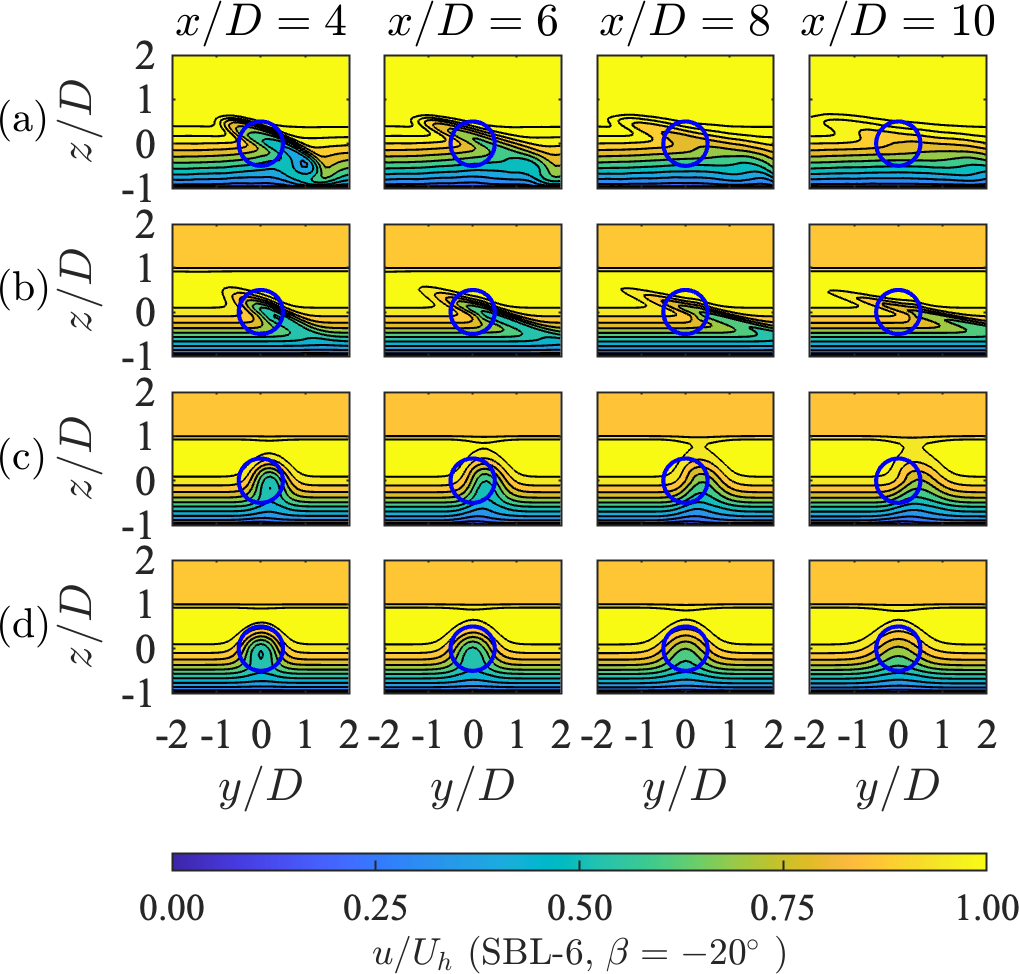}
    \caption{Contour plots of velocity field behind a  $\beta=-20^\circ$ yawed wind turbine in SBL-6 flow at streamwise locations $x/D=[4,6,8,10]$ from (a) LES, (b) veer-corrected vortex sheet-based wake model (M-1), (c) vortex sheet-based wake model without veer correction (M-2), and (d) Gaussian wake model (M-3).} \label{u_yaw-20_SBL6_M1M2M3}
\end{figure}

Figs. \ref{u_yaw0_SBL4_M1M2M3}, \ref{u_yaw20_SBL5_M1M2M3}, and \ref{u_yaw-20_SBL6_M1M2M3} present flow velocity contours in the wake region of an upstream turbine yawed at angles $\beta=0^\circ$, $\beta=20^\circ$, and $\beta=-20^\circ$, respectively, under SBL-4, SBL-5, and SBL-6 atmospheric conditions. In each figure, panel (a) shows contours from LES, while panels (b), (c), and (d) display the velocity field predictions from the M-1, M-2, and M-3 wake models, respectively.
LES contours clearly illustrate the deceleration of the flow velocity immediately behind the turbine, followed by a gradual increase in velocity at downstream locations due to wake recovery.
 In all Figs. \ref{u_yaw0_SBL4_M1M2M3}, \ref{u_yaw20_SBL5_M1M2M3}, and \ref{u_yaw-20_SBL6_M1M2M3}, the velocity distribution predicted by the M-1 wake model resembles the LES contours, particularly in capturing the tilt of the flow due to wind veer. In contrast, the M-2 and M-3 models do not capture the shearing effect caused by wind veer, resulting in significant deviations from the LES results. As a result, the M-1 wake model will more likely give better power loss predictions compared to the M-2 and M-3 models (see below). 

\subsubsection{Power losses}

In this section, we quantify the power loss caused by wake interaction by calculating $P(x_T, y_T)/P_0$ using Eq. \eqref{P_ratio} for an unyawed wind turbine positioned within the wake of an upstream turbine. The analysis considers different atmospheric conditions, including CNBL, SBL-4, 5, and 6, and for both unyawed/yawed conditions for the upstream turbine. Results are presented as spanwise profiles as a function of the spanwise displacement of the downstream turbine, $y_T$, at various discrete streamwise distances, $x_T$.

\begin{figure}[H]
    \centering
    \includegraphics[scale=1]{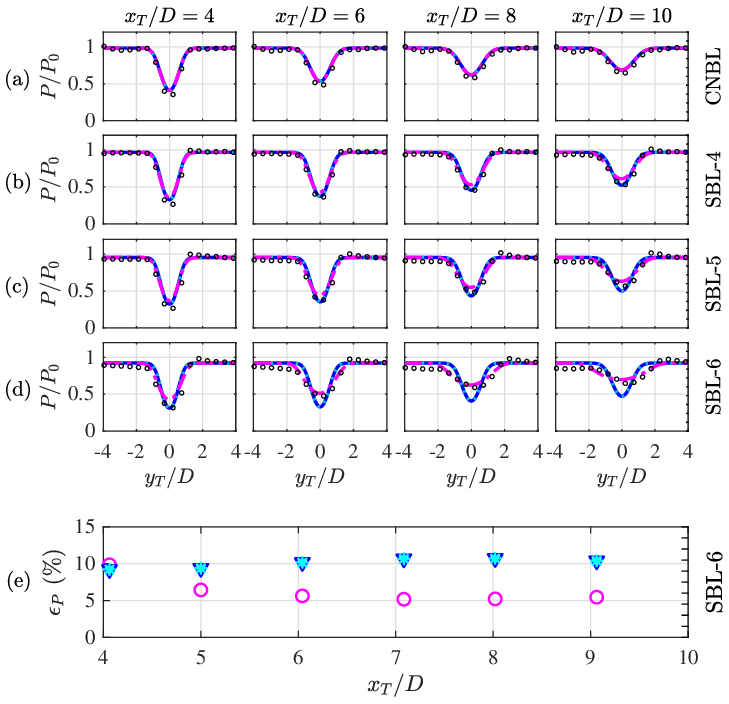}
    \caption{Power ratios $(P/P_0)$ (Eq. \ref{P_ratio}) of unyawed wind turbine placed behind the wake of an unyawed upstream turbine at locations $x_T/D=[4,6,8,10]$ and $y_T/D=[-4, 4]$. Plots of $P/P_0$ from LES ($\protect\blackcircopen$) compared against predictions from M-1 ($\protect\magentadashdotline$), M-2 ($\protect\blueline$), M-3 ($\protect\cyandotline$) in ABL flows: (a) CNBL, (b) SBL-4, (c) SBL-5, (d) SBL-6. Panel (e) shows the mean absolute percentage error ($\epsilon_P$ from Eq. \ref{eps_P}) of the power ratios between the LES and model predictions evaluated using M-1 ($\protect\magentacircopen$), M-2 ($\protect\bluetrianglemarker$), and M-3 (${\color{cyan}*}$) for SBL-6  case. 
    }
    \label{P_ratio_y_yaw0}
\end{figure}

\begin{figure}[H]
    \centering
    \includegraphics[scale=1]{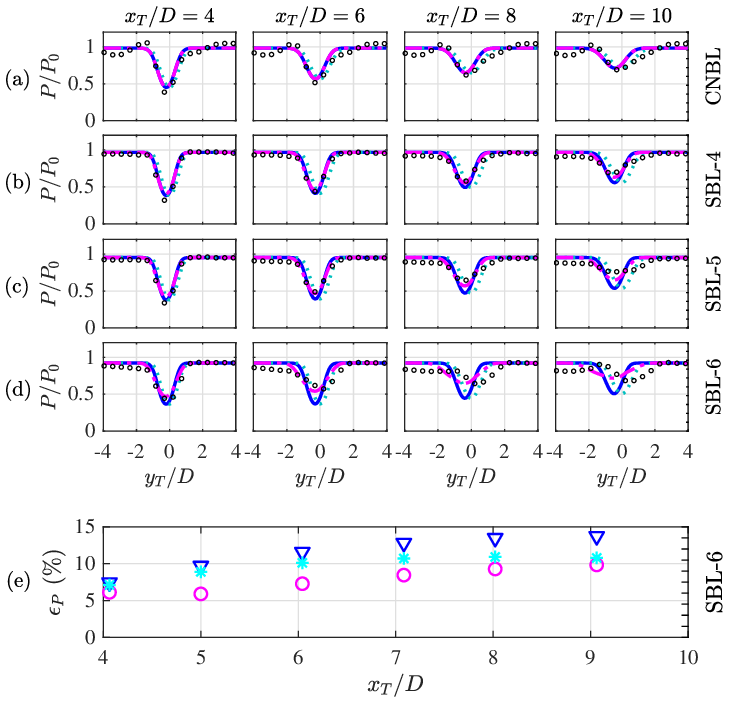}
    \caption{Power ratios $(P/P_0)$ (Eq. \ref{P_ratio}) of unyawed wind turbine placed behind the wake of a $\beta=20^\circ$ yawed upstream turbine. Lines and symbols are the same as in Fig. \ref{P_ratio_y_yaw0}.}\label{P_ratio_y_yaw20}
\end{figure}

\begin{figure}[H]
    \centering
    \includegraphics[scale=1]{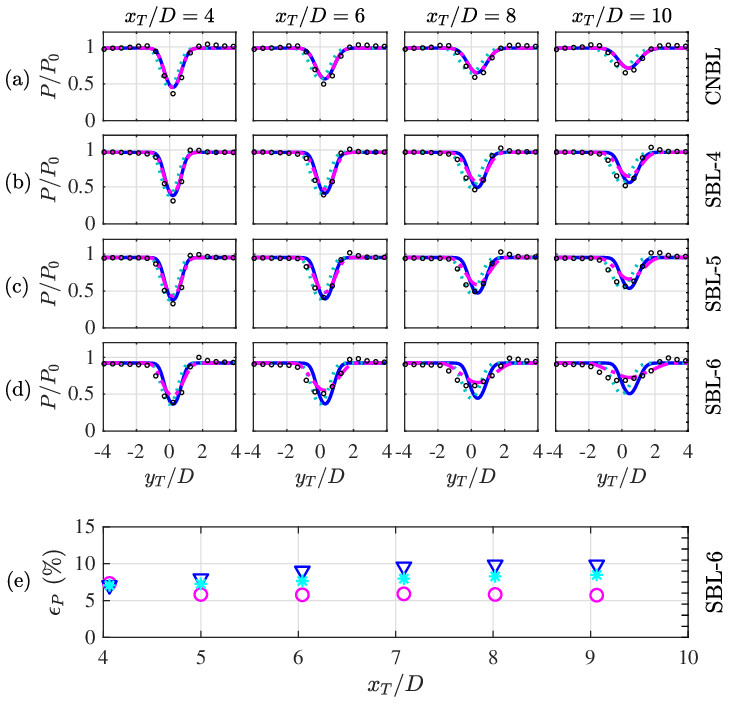}
    \caption{Power ratios $(P/P_0)$ (Eq. \ref{P_ratio}) of unyawed wind turbine placed behind the wake of a $\beta=-20^\circ$ yawed upstream turbine.
    Lines and symbols are the same as in Fig. \ref{P_ratio_y_yaw0}.}\label{P_ratio_y_yaw-20}
\end{figure}

Figures \ref{P_ratio_y_yaw0}, \ref{P_ratio_y_yaw20}, and \ref{P_ratio_y_yaw-20}  show power ratios across the range $y_T/D = [-4, 4]$ at  $x_T/D = [4, 6, 8, 10]$. In each figure, panel (a) is for the CNBL atmospheric flow, while panels (b), (c), and (d) show results for SBL-4, SBL-5, and SBL-6 conditions, respectively. The LES results for $P/P_0$ are represented by black circle markers ($\protect\blackcircopen$), whereas the predictions from the M-1, M-2, and M-3 models are shown by the dash-dotted magenta line ($\protect\magentadashdotline$), solid blue line ($\protect\blueline$), and dotted cyan line ($\protect\cyandotline$), respectively. Additionally, panel (e) in each figure presents the mean absolute percentage error ($\epsilon_P$) of the power ratios, comparing the wake model predictions against the LES results under SBL-6 atmospheric flow. The error $\epsilon_P(x_T)$ is evaluated as
\begin{align}
\epsilon_P(x_T) = \frac{1}{N^P_y} \sum_{j=1}^{N^P_y} 
\left| \frac{[P(x_T,y_{T,j})/P_0]_{\text{LES}} - [P(x_T,y_{T,j})/P_0]_{\text{model}}}{[P(x_T,y_{T,j})/P_0]_{\text{LES}}} \right| \times 100,\label{eps_P}
\end{align}
where $N_y^P$ is the number of spanwise grid points such that $-4<y_T/D<4$. The MAPE for $P/P_0$ using M-1, M-2, and M-3 are depicted by magenta circle markers ($\protect\magentacircopen$), blue triangle markers ($\protect\bluetrianglemarker$), and cyan asterisk markers (${\color{cyan}*}$), respectively.

In all these figures, as expected the downstream turbine experiences reduced power output ($P/P_0<1$) when located within the wake of the upstream turbine. 
Under neutral (CNBL) conditions for the unyawed turbine in Fig. \ref{P_ratio_y_yaw0}(a), the M-1, M-2, and M-3 wake models yield identical predictions for the $P/P_0$ ratio, as yaw-induced deflection and wind veer shearing are negligible in this scenario. Similarly, under yawed conditions in Figs. \ref{P_ratio_y_yaw20} and \ref{P_ratio_y_yaw-20}:(a), the minimum value of the power ratio shifts to the left for $\beta = 20^\circ$ and to the right for $\beta = -20^\circ$, corresponding to the yaw-induced deflection. Since the wind veer is weaker in the CNBL flow, the M-1 model predicts results similar to the M-2 model, while the M-3 model consistently places the minimum $P/P_0$ value at $y_T/D = 0$.

Under strong wind veer conditions, such as in SBL flows, panels (b)-(d) in Figs. \ref{P_ratio_y_yaw0}, \ref{P_ratio_y_yaw20}, and \ref{P_ratio_y_yaw-20} illustrate that the $P/P_0$ predictions using the M-1 wake model, which accounts for wind veer effects, more closely match the LES results compared to the M-2 and M-3 models.
To quantify the error, panel (e) in each figure presents the MAPE ($\epsilon_P$ from Eq. \ref{eps_P}) of $P/P_0$ between the LES results and model predictions for the SBL-6 atmospheric condition, where wind veer is most intense.

For all yaw angles, the MAPE for the M-1 model is consistently lower than that of the M-2 and M-3 models across various streamwise positions. In the unyawed case, as shown in Fig. \ref{P_ratio_y_yaw0}(e), the M-1 model's error is approximately 5\%, while the M-2 and M-3 models exhibit a higher error of about 10\%. For the $\beta=20^\circ$ yaw case, the M-1 model shows good agreement with LES results near the turbine, particularly at distances around $4D-6D$, where $\epsilon_P$ is around 5\%, although the error increases to 10\% further downstream. 
This rise in error can be attributed to the additional mixing caused by the tilted CVPs, which disrupt the wake, resulting in a more complex structure, as seen in Fig. \ref{fig:yaw20_du_LES_vs_model}(g) and Fig. \ref{fig:yaw20_du_LES_vs_model_xy}(c). These figures illustrate clear differences between the LES contours and the model predictions at larger downstream distances. Since the wake model does not account for the extra mixing induced by secondary flows or tilting of the vortices in the CVP, discrepancies in velocity deficit predictions occur, leading to higher errors in calculating $P/P_0$ at farther downstream locations for the $\beta=20^\circ$ case. Despite the increased error for M-1, the M-2 and M-3 models exhibit slightly larger errors, rendering the M-1 model more favorable also for these conditions. Similarly, in the $\beta=-20^\circ$ case, the M-1 model maintains an error of approximately 5\% between $4D$ and $10D$, while the M-2 and M-3 models show greater errors (almost twice as large). Overall, this analysis demonstrates that the new analytical wake model (M-1), which includes veer effects, provides improved predictions of power loss due to wake interactions compared to some examples of earlier models that do not aim to include these more complex effects. 

\section{Conclusion} \label{sec:conclusion}
In this study, we introduced a new analytical wake model designed to describe the wake structures behind wind turbines in ABL flows. Specifically, we focused on modeling wakes generated by turbines, both with and without yaw, operating in CNBL and SBL atmospheric conditions. This new model builds upon the vortex sheet-based wake model by \textcite{bastankhah_et_al_2022}, originally developed for modeling yawed turbine wakes in neutral atmospheric flows. We extend this model to account for the effects of wind veer and thermal stratification. Following the approach of \textcite{Narasimhan_et_al_2022}, the veer-induced spanwise deflection $(y_{c,\text{veer}}=x V(z)/U(z))$ is additively combined with the yaw-induced wake deflection $(\hat{y}c(\hat{t})R\sqrt{A_*})$ in Eq. \eqref{yc_xz}. The curled wake structure induced by yawing is captured by the polar-angle dependent wake width given by Eq. \eqref{sigma_width}. 
The veer-induced deflection $y_{c,\text{veer}}$ is computed using the analytical ABL velocity profiles from the new coupled Ekman-surface layer ABL model proposed by \textcite{Narasimhan_et_al_BLM_2024}.

The wake model was validated using data from LES, including conventionally neutral and stable ABL flows and for yawed and unyawed turbines. We demonstrated that the predictions of the velocity deficit distribution and its decay from the refined veer-corrected curled wake model show good agreement with LES results where we discussed that the MAPE of the flow velocities in the far downstream regions for $x/D>3-5$ is less than 2\%.  Additionally, we demonstrated that this enhanced wake model offers improved predictions compared to existing models for predicting the power loss of an unyawed wind turbine positioned in the wake of a preceding yawed or unyawed wind turbine. The errors in the power loss predictions using the new wake model ranged between 5\%-10\% across the different yaw and ABL conditions, in all cases markedly better than models that omitted veer, stratification, and curling effects. We can conclude that the proposed extended analytical wake model is capable of describing wakes behind turbines, yawed or unyawed, across a relatively broad range of atmospheric stability conditions.

Future research should aim to extend the ABL model to describe convective boundary layers and account for unsteady effects, such as those observed during a diurnal cycle. Further refinements of the ABL model, including incorporating additional factors like momentum exchanges with surface canopies \citep{patton2016atmospheric}, are also of significant interest. Moreover, it bears recalling that the wake model developed in this study is designed to predict wake characteristics for individual wind turbines within ABL flows. Future efforts should also aim to extend this modeling approach to predict wake structures for entire wind farms, accounting for both onshore and offshore wind farms in marine ABL flows \citep{sullivan_et_al_2014}. This can be achieved by 
using wake superposition methods \citep{Jensen_1983,stevens_meneveau_2017} and establishing generalizations of the coupled Ekman-surface layer ABL model for fully developed wind turbine array boundary layer (FD-WTABL) mean velocity distributions \citep{calaf_et_al_2010,meneveau_2012}.

{\noindent \bf Acknowledgements:}  
This research was supported by the National Science Foundation (NSF) through grants CBET-1949778 and CMMI-2034111. We also acknowledge the use of Cheyenne High-Performance Computing resource ({doi:10.5065/D6RX99HX}), provided by the NCAR's CISL, 
sponsored by the NSF. Additionally, we are grateful for the resources from the Advanced Research Computing at Hopkins (ARCH) core facility (rockfish.jhu.edu), supported by NSF grant 
OAC-1920103. 

We would like to dedicate this work to the memory of two notable scholars who have sadly passed away this year:  Prof. Harihar Rajaram, to whom we are grateful for his insightful discussions and for serving as an examiner for Ghanesh Narasimhan's PhD dissertation defense, and Prof. Dries Allaerts to whom we are grateful for his encouraging feedback on this work during the TORQUE 2024 conference, and for his enduring contributions to the community.

\nocite{*}

\bibliography{wake_model_ABL_JRSE_2024}

\end{document}